\documentstyle[12pt]{%
article}\newlength\Cscr\newlength
\Csave\newlength\Ctenthex\newlength
\CFxsize\newlength\CFxsizeps\newlength\CFsizemakebox\newlength
\CFleftcrop\newlength\CFrightcrop\newlength\CZtbldist\newlength
\CZfigdist
\newlength\CGDnum\newlength\CGDtext\newcounter{Cscr}\newcounter{%
Csave}\newcounter{CBcit}\newcounter{CElett}%
\newcounter{CBauthornu%
m}\newcounter{Ceqindent}\newcounter{CBtnc}
\newcounter{CBtntc}\newcounter{CEht}%
\newcounter{CbsA}\newcounter{CbsB}\newcounter
{CbsC}\newcounter{CbsD}\hoffset\Cscr\voffset
\Cscr\protect
\begin{document}\renewcommand\theequation{\arabic{e%
quation}}\renewcommand\thetable{\arabic{table}}\renewcommand
\thefigure{\arabic{figure}}\renewcommand\thesection{\Roman{secti%
on}}\renewcommand\thesubsection{\Alph{subsection}}\renewcommand
\thesubsubsection{\arabic{subsubsection}}\setcounter{CEht}{10}%
\setcounter{CbsA}{1}\setcounter{CbsB}{1}\setcounter{CbsC}{1}%
\setcounter{CbsD}{1}\hfill MG--P--01/051\par\hfill To appear in 
the \mbox{}\protect\/{\protect\em American Journal of Physics%
\protect\/}\par{\centering{\protect\mbox{}}\\*[\baselineskip]{%
\large\bf The Thomas rotation}\\*}\addtocounter{CBtntc}{1}%
\addtocounter{CBtntc}{1}\addtocounter{CBtntc}{1}\addtocounter{CB%
tntc}{1}{\centering{\protect\mbox{}}\\John P.~Costella,$^{%
\fnsymbol{CBtnc}}$\addtocounter{CBtnc}{1}$^{\fnsymbol{CBtnc}}$%
\addtocounter{CBtnc}{1}\ Bruce H.~J.~McKellar,$^{\fnsymbol{CBtnc%
}}$\addtocounter{CBtnc}{1}\ and Andrew A.~Rawlinson$^{\fnsymbol{%
CBtnc}}$\addtocounter{CBtnc}{1}\\*}{\centering{\small\mbox{}%
\protect\/{\protect\em School of Physics, The University of Melb%
ourne, Parkville, Victoria 3052, Australia\protect\/}}\\}%
\addtocounter{CBtnc}{1}\addtocounter{CBtntc}{1}{\centering{%
\protect\mbox{}}\\Gerard J.~Stephenson, Jr.$^{\fnsymbol{CBtnc}}$%
\addtocounter{CBtnc}{1}\\*}{\centering{\small\mbox{}\protect\/{%
\protect\em Department of Physics and Astronomy, University of N%
ew Mexico, Albuquerque, NM 87131\protect\/}}\\}{\centering{%
\protect\mbox{}}\\(February\ 20, 2001)\\} \par\vspace
\baselineskip\begin{center}{\small\bf Abstract}\end{center}%
\vspace{-1.25ex}\vspace{-0.75\baselineskip}\par\setlength{\Csave
}{\parskip}\begin{quote}\setlength{\parskip}{0.5\baselineskip}%
\small\noindent We review why the \mbox{}\protect\/{\protect\em
Thomas rotation\protect\/} is a crucial facet of special relativ%
ity, that is just as fundamental, and just as ``unintuitive'' an%
d ``paradoxical''\mbox{$\!$}, as such traditional effects as len%
gth contraction, time dilation, and the ambiguity of simultaneit%
y. We show how this phenomenon can be quite naturally introduced 
and investigated in the context of a typical introductory course 
on special relativity, in a way that is appropriate for, and com%
pletely accessible to, undergraduate students. We also demonstra%
te, in a more advanced section aimed at the graduate student stu%
dying the Dirac equation and relativistic quantum field theory, 
that careful consideration of the Thomas rotation will become vi%
tal as modern experiments in particle physics continue to move f%
rom unpolarized to polarized cross-sections. \end{quote}%
\setlength{\parskip}{\Csave}\par\refstepcounter{section}\vspace{%
1.5\baselineskip}\par{\centering\bf\thesection. Introduction\\*[%
0.5\baselineskip]}\protect\indent\label{sect:Intro}Recently, a n%
umber of the current authors have reviewed how aspects of relati%
vistic quantum mechanics can be appreciated from the point of vi%
ew of relativistic \mbox{}\protect\/{\protect\em classical%
\protect\/} mechanics. In Ref.~\ref{cit:Costella1995}, the Foldy%
--Wout\-huysen transformation was reviewed, where it was emphasi%
zed that many of the operators of the Dirac equation become, aft%
er transformation, completely recognizable from the point of vie%
w of classical physics. In Ref.~\ref{cit:Costella1997}, the Feyn%
man--Stueckelberg formulation of antiparticles was reviewed, ent%
irely within the domain of classical mechanics, and it was empha%
sized that one can make good sense of antiparticle motion withou%
t needing to resort to quantum mechanical arguments.\par In exte%
nding these ideas to the domain of quantum field theory, we have 
found that there is a third aspect of classical relativistic mec%
hanics that is of crucial theoretical and practical importance, 
but which rates scarcely a mention in most textbooks on special 
relativity: the \mbox{}\protect\/{\protect\em Thomas rotation%
\protect\/}. (In the case of a continuous evolution of infinites%
imal rotations, this effect is usually referred to as the Thomas 
\mbox{}\protect\/{\protect\em precession\protect\/}; but we are 
here mainly concerned with the more general case of a single, 
\mbox{}\protect\/{\protect\em finite\protect\/} rotation.) Histo%
rically, the relative obscurity of this effect can, perhaps, be 
traced to the fact that the special theory of relativity was two 
decades old before Thomas made his discovery. Pais's summary of 
events$^{\ref{cit:Pais1982}}$ is instructive: \begin{quote} 
\small Twenty years later [after his seminal 1905 paper on speci%
al relativity], Einstein heard something about the Lorentz group 
that greatly surprised him. It happened while he was in Leiden. 
In October 1925 George Eugene Uhlenbeck and Samuel Goudsmit had 
discovered the spin of the electron and thereby explained the oc%
currence of the alkali doublets, but for a brief period it appea%
red that the magnitude of the doublet splitting did not come out 
correctly. Then Llewellyn Thomas supplied the missing factor, 2, 
now known as the Thomas factor. Uhlenbeck told me that he did no%
t understand a word of Thomas's work when it first came out. `I 
remember that, when I first heard about it, it seemed unbelievab%
le that a relativistic effect could give a factor of 2 instead o%
f something of order $v/c$~\ldots. Even the cognoscenti of the r%
elativity theory (Einstein included!)\ were quite surprised.' At 
the heart of the Thomas precession lies the fact that a Lorentz 
transformation with velocity \mbox{$\protect\displaystyle{%
\protect\mbox{\protect\boldmath{$v$}}}_{1\!}$} followed by a sec%
ond one with velocity \mbox{$\protect\displaystyle{\protect\mbox
{\protect\boldmath{$v$}}}_{2}$} in a different direction does no%
t lead to the same inertial frame as one single Lorentz transfor%
mation with the velocity \mbox{$\protect\displaystyle{\protect
\mbox{\protect\boldmath{$v$}}}_{1\!}+{\protect\mbox{\protect
\boldmath{$v$}}}_2$}. (It took Pauli a few weeks before he grasp%
ed Thomas's point.) \end{quote} It seems remarkable---but, accor%
ding to the above account, undeniable---that neither Einstein no%
r Pauli came across the Thomas rotation before 1925. However, th%
e effect we now call the Thomas rotation was known before Thomas%
's paper. The early history has been traced by Ungar.$^{\ref{cit%
:Ungar1991}}$\par Now, most textbooks on special relativity foll%
ow the extraordinarily clear exposition of the theory given by E%
instein in his seminal paper. Unfortunately, this has meant that 
little or no attention has usually been given to the ``Thomas ef%
fect''\mbox{$\!$}, which has generally been relegated to a brief 
mention in textbooks on quantum mechanics and atomic structure. 
As far as we are aware, the best treatment of the Thomas \mbox{}%
\protect\/{\protect\em precession\protect\/} in a textbook still 
in print is arguably that contained in Jackson's book on classic%
al electrodynamics.$^{\ref{cit:Jackson1999}}$ A similar discussi%
on is given in Goldstein,$^{\ref{cit:Goldstein1980}}$ who emphas%
izes the complexity of the general calculations. This complexity 
is inhibiting to both the writers and the readers of the textboo%
ks.\par That the Thomas rotation, or precession, still puzzles s%
tudents and their teachers can be discerned from the pages of th%
is journal. In Question~\#57, MacKeown$^{\ref{cit:MacKeown1997}}%
$ asks ``\ldots\ is said to introduce a velocity independent con%
stant factor. Can any simple, convincing, argument be given for 
this?'' Ungar$^{\ref{cit:Ungar1991}}$ and Goedecke$^{\ref{cit:Go%
edecke1978}}$ have countered the complexity by introducing new f%
ormalisms, a ``weakly associative-commutative groupoid'' by Unga%
r, and the tetrad formalism by Goedecke. While they offer useful 
insights, and emphasise the Thomas rotation, they are not well s%
uited to the introductory course. Muller$^{\ref{cit:Muller1992}}%
$ (in the Appendix) and Philpott$^{\ref{cit:Philpott1996}}$ (to 
introduce the main point of his paper) give derivations of the T%
homas precession which are related to the present one. But we be%
lieve that the straightforward treatment below, and its emphasis 
on Thomas \mbox{}\protect\/{\protect\em rotation\protect\/}, off%
ers conceptual and pedagogical advantages which make it suitable 
to an introductory course.\par In this paper, we show how an ins%
tructive, elementary, and intriguing discussion of the Thomas ro%
tation can be ``grafted on'' to any standard introductory course 
on special relativity. As prerequisite we assume nothing more th%
an the standard expression for a Lorentz boost along the $x$-axi%
s of a system of co\-ordinates. For simplicity, we also make use 
of the energy--momentum four-vector, as well as matrix multiplic%
ation, although such references could be deleted if thought nece%
ssary, albeit at the expense of rendering the algebra a little l%
ess transparent. (The widespread availability of calculators and 
computer programs capable of matrix multiplication means that th%
e complexities of the following calculations can be drastically 
\mbox{}\protect\/{\protect\em minimized\protect\/} by the use of 
matrix notation---leaving more time for the contemplation of the 
physical results.) These preliminaries are covered in \mbox{Sec.%
 ~$\:\!\!$}\protect\ref{sect:Prelim}.\par In \mbox{Sec.~$\:\!\!$%
}\protect\ref{sect:Paradox}, we show how these simple building b%
locks can be put together to create a sequence of intriguing and 
completely counterintuitive ``paradoxes''\mbox{$\!$}. This mater%
ial could be presented almost verbatim in any introductory cours%
e on special relativity.\par\mbox{Sec.~$\:\!\!$}\protect\ref{sec%
t:Solution} provides full solutions and explanations of these el%
ementary Thomas rotation ``paradoxes''\mbox{$\!$}, and general e%
xpressions are derived for the Thomas rotation in arbitrary case%
s.\par In \mbox{Sec.~$\:\!\!$}\protect\ref{sect:Scatter}, we pro%
vide a further ``paradox'' in the context of the polarization pr%
operties of the scattering of a Dirac particle. This example is 
more advanced, in that it presumes familiarity with at least an 
introductory level of quantum field theory; and thus it would no%
t usually be appropriate for an introductory course on special r%
elativity. On the other hand, this example is arguably much more 
\mbox{}\protect\/{\protect\em practically\protect\/} important t%
han the others, in that it shows that real-life calculations of 
scattering cross-sections can be completely erroneous if due reg%
ard is not taken of this subtle facet of relativistic kinematics%
. \mbox{Sec.~$\:\!\!$}\protect\ref{sect:ScattSol} provides a ful%
l solution of this ``paradox''\mbox{$\!$}.\par Finally, \mbox{Se%
c.~$\:\!\!$}\protect\ref{sect:Conclude} summarizes our conclusio%
ns.\par\refstepcounter{section}\vspace{1.5\baselineskip}\par{%
\centering\bf\thesection. Preliminaries\\*[0.5\baselineskip]}%
\protect\indent\label{sect:Prelim}In this section we review thos%
e features of special relativity which we would assume to have b%
een taught in an introductory course before the discussion of Th%
omas rotation, to set the scene and to establish our notation.%
\par Throughout this paper, we shall use a ``naturalized'' set o%
f units, in which \mbox{$\protect\displaystyle c=1$}. To convert 
any expression to SI~units, one need simply replace $t$~by~$ct$, 
${\protect\mbox{\protect\boldmath{$v$}}}$~by~${\protect\mbox{%
\protect\boldmath{$v$}}}/c$, $E$~by~\mbox{$\protect\displaystyle
E/{c}{}^{\raisebox{-0.25ex}{$\scriptstyle\:\!2$}}$}\mbox{$\!$}, 
and ${\protect\mbox{\protect\boldmath{$p$}}}$~by~${\protect\mbox
{\protect\boldmath{$p$}}}/c$. (Boldface denotes a three-vector.) 
In \mbox{Secs.~$\:\!\!$}\protect\ref{sect:Scatter} and~\protect
\ref{sect:ScattSol} we shall also use units in which \mbox{$%
\protect\displaystyle\hbar=1$}.\par The Lorentz transformation f%
rom a frame $S$ with co\-ordinates \mbox{$\protect\displaystyle(%
t,x,y,z)$}, to a frame $S'$\mbox{$\!$}, moving with respect to $%
S$ with a velocity $v$ along the $x$-axis, in which the co\-ordi%
nates are \mbox{$\protect\displaystyle(t'\!,x'\!,y'\!,z')$}, is 
\setcounter{Ceqindent}{0}\protect\begin{eqnarray}\protect\left.%
\protect\begin{array}{rcl}\protect\displaystyle\hspace{-1.3ex}&%
\protect\displaystyle t^{\prime\!}=\gamma(t-vx),\hspace{8ex}x^{%
\prime\!}=\gamma(x-vt),\hspace{8ex}y^{\prime\!}=y,\hspace{8ex}z^%
{\prime\!}=z,\setlength{\Cscr}{\value{CEht}\Ctenthex}\addtolength
{\Cscr}{-1.0ex}\protect\raisebox{0ex}[\value{CEht}\Ctenthex][%
\Cscr]{}\protect\end{array}\protect\right.\protect\label{eq:Prel%
im-BxCmpts}\protect\end{eqnarray}\setcounter{CEht}{10}where 
\setcounter{Ceqindent}{0}\protect\begin{eqnarray}\protect\left.%
\protect\begin{array}{rcl}\protect\displaystyle\hspace{-1.3ex}&%
\protect\displaystyle\gamma\equiv\mbox{$\protect\displaystyle
\protect\frac{1}{\sqrt{\raisebox{-0.15ex}[1.5ex][0ex]{$1-{v}{}^{ 
\raisebox{-0.25ex}{$\scriptstyle2$}}$}}}$}.\setlength{\Cscr}{%
\value{CEht}\Ctenthex}\addtolength{\Cscr}{-1.0ex}\protect
\raisebox{0ex}[\value{CEht}\Ctenthex][\Cscr]{}\protect\end{array%
}\protect\right.\protect\label{eq:Prelim-Gamma}\protect\end{eqna%
rray}\setcounter{CEht}{10}This transformation, written in matrix 
notation, is \setcounter{Ceqindent}{0}\protect\begin{eqnarray}%
\protect\left.\protect\begin{array}{rcl}\protect\displaystyle
\hspace{-1.3ex}&\protect\displaystyle\left(\begin{array}{c}t'\\x%
'\\y'\\z'\end{array}\right)=\left(\begin{array}{cccc}\gamma&\!\!%
\!-\gamma v\!&\,0\,&\,0\,\\\!\!\!-\gamma v\!&\gamma&0&0\\0&0&1&0%
\\0&0&0&1\end{array}\right)\!\left(\begin{array}{c}t\\x\\y\\z\end
{array}\right)\mbox{$\!$}.\setlength{\Cscr}{\value{CEht}\Ctenthex
}\addtolength{\Cscr}{-1.0ex}\protect\raisebox{0ex}[\value{CEht}%
\Ctenthex][\Cscr]{}\protect\end{array}\protect\right.\protect
\label{eq:Prelim-BxMatrix}\protect\end{eqnarray}\setcounter{CEht%
}{10}We shall denote the matrix that effects this boost by veloc%
ity $v$ in the $x$-direction as $B_x(v)$: \setcounter{Ceqindent}%
{0}\protect\begin{eqnarray}\protect\left.\protect\begin{array}{r%
cl}\protect\displaystyle\hspace{-1.3ex}&\protect\displaystyle B_%
x(v)\equiv\left(\begin{array}{cccc}\gamma&\!\!\!-\gamma v\!&\,0%
\,&\,0\,\\\!\!\!-\gamma v\!&\gamma&0&0\\0&0&1&0\\0&0&0&1\end{arr%
ay}\right)\mbox{$\!$}.\setlength{\Cscr}{\value{CEht}\Ctenthex}%
\addtolength{\Cscr}{-1.0ex}\protect\raisebox{0ex}[\value{CEht}%
\Ctenthex][\Cscr]{}\protect\end{array}\protect\right.\protect
\label{eq:Prelim-Bx}\protect\end{eqnarray}\setcounter{CEht}{10}C%
learly, a boost by velocity $v$ in the $y$-direction or in the $%
z$-direction would be effected by \setcounter{Ceqindent}{0}%
\protect\begin{eqnarray}\protect\left.\protect\begin{array}{rcl}%
\protect\displaystyle\hspace{-1.3ex}&\protect\displaystyle B_y(v%
)\equiv\left(\begin{array}{cccc}\gamma&\,0\,&\!\!\!-\gamma v\!&%
\,0\,\\0&1&0&0\\\!\!\!-\gamma v\!&0&\gamma&0\\0&0&0&1\end{array}%
\right)\mbox{$\!$},\hspace{10ex}B_z(v)\equiv\left(\begin{array}{%
cccc}\gamma&\,0\,&\,0\,&\!\!\!-\gamma v\!\\0&1&0&0\\0&0&1&0\\\!%
\!\!-\gamma v\!&0&0&\gamma\end{array}\right)\mbox{$\!$}.%
\setlength{\Cscr}{\value{CEht}\Ctenthex}\addtolength{\Cscr}{-1.0%
ex}\protect\raisebox{0ex}[\value{CEht}\Ctenthex][\Cscr]{}\protect
\end{array}\protect\right.\protect\label{eq:Prelim-ByBz}\protect
\end{eqnarray}\setcounter{CEht}{10}The \mbox{}\protect\/{\protect
\em energy--momentum four-vector\protect\/}, \setcounter{Ceqinde%
nt}{0}\protect\begin{eqnarray}\protect\left.\protect\begin{array%
}{rcl}\protect\displaystyle\hspace{-1.3ex}&\protect\displaystyle
p^{\:\!\mu\!}=\left(\begin{array}{c}E\\p^{\:\!x\!}\\p^{\:\!y\!}%
\\p^{\:\!z\!}\end{array}\right)\mbox{$\!$},\setlength{\Cscr}{%
\value{CEht}\Ctenthex}\addtolength{\Cscr}{-1.0ex}\protect
\raisebox{0ex}[\value{CEht}\Ctenthex][\Cscr]{}\protect\end{array%
}\protect\right.\protect\label{eq:Prelim-Pmu}\protect\end{eqnarr%
ay}\setcounter{CEht}{10}shall play a key role\ in our analysis. 
A particle of mass $m$, at rest in a system of co\-ordinates, ha%
s \mbox{$\protect\displaystyle E=m$} and \mbox{$\protect
\displaystyle{\protect\mbox{\protect\boldmath{$p$}}}=\mbox{\bf0}%
$}. If we boost our system of co\-ordinates by the velocity $-v$ 
in the $x$-direction (so that, relative to this new system of co%
\-ordinates, the particle has velocity $+v$ in the $x$-direction%
), then the application of $B_x(-v)$ yields \setcounter{Ceqinden%
t}{0}\protect\begin{eqnarray}\hspace{-1.3ex}&\displaystyle\left(%
\begin{array}{c}E\\p^{\:\!x\!}\\p^{\:\!y\!}\\p^{\:\!z\!}\end{arr%
ay}\right)=\left(\begin{array}{cccc}\gamma&\!\gamma v\!&\,0\,&\,%
0\,\\\!\gamma v\!&\gamma&0&0\\0&0&1&0\\0&0&0&1\end{array}\right)%
\!\left(\begin{array}{c}m\\0\\0\\0\end{array}\right)=\left(\begin
{array}{c}m\gamma\\m\gamma v\\0\\0\end{array}\right)\mbox{$\!$}.%
\protect\nonumber\setlength{\Cscr}{\value{CEht}\Ctenthex}%
\addtolength{\Cscr}{-1.0ex}\protect\raisebox{0ex}[\value{CEht}%
\Ctenthex][\Cscr]{}\protect\end{eqnarray}\setcounter{CEht}{10}Th%
is implies that the $x$-velocity of the particle can be ``extrac%
ted'' from the components \mbox{$\protect\displaystyle p^{\:\!\mu
\!}$} of its four-momentum by computing the ratio \mbox{$\protect
\displaystyle p^{\:\!x\!}/E$}. Since the $x$-direction is arbitr%
ary, the immediate generalization to a particle moving with velo%
city ${\protect\mbox{\protect\boldmath{$v$}}}$ in any direction 
is \setcounter{Ceqindent}{0}\protect\begin{eqnarray}\protect\left
.\protect\begin{array}{rcl}\protect\displaystyle\hspace{-1.3ex}&%
\protect\displaystyle{\protect\mbox{\protect\boldmath{$v$}}}=%
\mbox{$\protect\displaystyle\protect\frac{{\protect\mbox{\protect
\boldmath{$p$}}}}{E}$}.\setlength{\Cscr}{\value{CEht}\Ctenthex}%
\addtolength{\Cscr}{-1.0ex}\protect\raisebox{0ex}[\value{CEht}%
\Ctenthex][\Cscr]{}\protect\end{array}\protect\right.\protect
\label{eq:Prelim-VfromP}\protect\end{eqnarray}\setcounter{CEht}{%
10}To obtain the law for the composition of two velocities $v_1$ 
and $v_2$ in the same direction, we may simply apply $B_x(-v_1)$ 
followed by $B_x(-v_2)$ to a particle at rest: \setcounter{Ceqin%
dent}{0}\protect\begin{eqnarray}\protect\left.\protect\begin{arr%
ay}{rcl}\protect\displaystyle\hspace{-1.3ex}&\protect
\displaystyle B_x(-v_2)^{\,}B_x(-v_1)\!\left(\begin{array}{c}m\\%
0\\0\\0\end{array}\right)=\left(\begin{array}{c}m\gamma_1\gamma_%
2(1+v_1v_2)\\m\gamma_1\gamma_2(v_{1\!}+v_2)\\0\\0\end{array}%
\right)\mbox{$\!$}.\setlength{\Cscr}{\value{CEht}\Ctenthex}%
\addtolength{\Cscr}{-1.0ex}\protect\raisebox{0ex}[\value{CEht}%
\Ctenthex][\Cscr]{}\protect\end{array}\protect\right.\protect
\label{eq:Prelim-Collinear}\protect\end{eqnarray}\setcounter{CEh%
t}{10}(In this and subsequent equations we shall take it to be u%
nderstood that the four components of the column vector refer to 
the components of the four-momentum.) On using (\protect\ref{eq:%
Prelim-VfromP}), \mbox{Eq.~\mbox{$\protect\displaystyle\!$}}(%
\protect\ref{eq:Prelim-Collinear}) yields \setcounter{Ceqindent}%
{0}\protect\begin{eqnarray}\protect\left.\protect\begin{array}{r%
cl}\protect\displaystyle\hspace{-1.3ex}&\protect\displaystyle v_%
x=\mbox{$\protect\displaystyle\protect\frac{v_{1\!}+v_2}{1+v_1v_%
2}$},\setlength{\Cscr}{\value{CEht}\Ctenthex}\addtolength{\Cscr}%
{-1.0ex}\protect\raisebox{0ex}[\value{CEht}\Ctenthex][\Cscr]{}%
\protect\end{array}\protect\right.\protect\label{eq:Prelim-AddVS%
td}\protect\end{eqnarray}\setcounter{CEht}{10}namely, the standa%
rd result. It will be noted that the same result would have been 
obtained if we had applied $B_x(-v_2)$ first and $B_x(-v_1)$ sec%
ond: boosts in the same direction commute.\par\refstepcounter{se%
ction}\vspace{1.5\baselineskip}\par{\centering\bf\thesection. So%
me elementary Thomas rotation ``paradoxes''\\*[0.5\baselineskip]%
}\protect\indent\label{sect:Paradox}Let us now apply the results 
obtained in \mbox{Sec.~$\:\!\!$}\protect\ref{sect:Prelim} to som%
e hypothetical maneuvers of the \mbox{}\protect\/{\protect\em US%
S~Enterprise\protect\/} under impulse power. In the following, t%
he system of co\-ordinates being considered is that of an observ%
er on board the bridge of the \mbox{}\protect\/{\protect\em Ente%
rprise\protect\/}.\par Let us assume that the \mbox{}\protect\/{%
\protect\em Enterprise\protect\/}\ begins at rest relative to so%
me particular star. We ignore gravitational effects, so that if 
the \mbox{}\protect\/{\protect\em Enterprise\protect\/}\ were to 
not fire any thrusters, then it would remain at rest relative to 
the star.\par Let us now apply a boost to the \mbox{}\protect\/{%
\protect\em Enterprise\protect\/}\ by some velocity $v_0$ in the 
$x$-direction, and follow it by a boost by the velocity $-v_0$, 
again in the $x$-direction. We expect that the net effect on the 
velocity of the \mbox{}\protect\/{\protect\em Enterprise\protect
\/}\ would be zero: it would move in the $x$-direction during th%
e maneuver\ (by what distance is of no interest to us here), but 
at the end of the maneuver\ it would again be at rest relative t%
o the star. We can confirm this by considering the effect of app%
lying $B_x(v_0)$ and then $B_x(-v_0)$ on, say, the components of 
the four-momentum of the star, as observed from the \mbox{}%
\protect\/{\protect\em Enterprise\protect\/}. If the star has th%
e mass~$m$, then a straightforward calculation verifies that 
\setcounter{Ceqindent}{0}\protect\begin{eqnarray}\hspace{-1.3ex}%
&\displaystyle B_x(-v_0)^{\,}B_x(v_0)\!\left(\begin{array}{c}m\\%
0\\0\\0\end{array}\right)=\left(\begin{array}{c}m\\0\\0\\0\end{a%
rray}\right)\!,\protect\nonumber\setlength{\Cscr}{\value{CEht}%
\Ctenthex}\addtolength{\Cscr}{-1.0ex}\protect\raisebox{0ex}[%
\value{CEht}\Ctenthex][\Cscr]{}\protect\end{eqnarray}\setcounter
{CEht}{10}and indeed \mbox{$\protect\displaystyle B_x(-v_0)^{\,}%
B_x(v_0)=I$}, where $I$ is the identity matrix. (In performing t%
hese calculations, and those that follow, it is useful to replac%
e even powers of $v$, wherever they occur, by means of the ident%
ity \setcounter{Ceqindent}{0}\protect\begin{eqnarray}\protect
\left.\protect\begin{array}{rcl}\protect\displaystyle\hspace{-1.%
3ex}&\protect\displaystyle v^2\equiv1-\mbox{$\protect
\displaystyle\protect\frac{1}{\gamma^{2\!}}$},\setlength{\Cscr}{%
\value{CEht}\Ctenthex}\addtolength{\Cscr}{-1.0ex}\protect
\raisebox{0ex}[\value{CEht}\Ctenthex][\Cscr]{}\protect\end{array%
}\protect\right.\protect\label{eq:Paradox-VGamma}\protect\end{eq%
narray}\setcounter{CEht}{10}which can be derived from the defini%
tion (\protect\ref{eq:Prelim-Gamma}).) We can perform the same m%
aneuver\ in the $y$-direction: namely, \setcounter{Ceqindent}{0}%
\protect\begin{eqnarray}\hspace{-1.3ex}&\displaystyle B_y(-v_0)^%
{\,}B_y(v_0)\!\left(\begin{array}{c}m\\0\\0\\0\end{array}\right)%
=\left(\begin{array}{c}m\\0\\0\\0\end{array}\right)\!,\protect
\nonumber\setlength{\Cscr}{\value{CEht}\Ctenthex}\addtolength{%
\Cscr}{-1.0ex}\protect\raisebox{0ex}[\value{CEht}\Ctenthex][\Cscr
]{}\protect\end{eqnarray}\setcounter{CEht}{10}and \mbox{$\protect
\displaystyle B_y(-v_0)^{\,}B_y(v_0)=I$}, again as expected.\par
Having thus verified the action of our ``thrusters'' in the $x$- 
and $y$-directions, by means of these four boosts, let us now tr%
y another test maneuver, by mixing the order of these boosts. Na%
mely, let us apply $B_x(v_0)$, followed by $B_y(v_0)$, then $B_x%
(-v_0)$, and finally $B_y(-v_0)$. Again, we expect that the star 
will be at rest, relative to the \mbox{}\protect\/{\protect\em E%
nterprise\protect\/}, at the end of the maneuver. However, if we 
perform the calculations, then (after some algebra) we find 
\setcounter{Ceqindent}{0}\protect\begin{eqnarray}\protect\left.%
\protect\begin{array}{rcl}\protect\displaystyle\hspace{-1.3ex}&%
\protect\displaystyle B_y(-v_0)^{\,}B_x(-v_0)^{\,}B_y(v_0)^{\,}B%
_x(v_0)\!\left(\begin{array}{c}m\\0\\0\\0\end{array}\right)=\left
(\begin{array}{c}m+m(\gamma_0+1)(\gamma_0-1)^3\\m\gamma_0^2(%
\gamma_{0\!}^{\protect\mbox{}}-1)v_0\\-m\gamma_0^{\protect\mbox{%
}}(\gamma_{0\!}^{\protect\mbox{}}-1)(-\gamma_0^2+\gamma_{0\!}^{%
\protect\mbox{}}+1)v_0\\0\end{array}\right)\mbox{$\!$}.\setlength
{\Cscr}{\value{CEht}\Ctenthex}\addtolength{\Cscr}{-1.0ex}\protect
\raisebox{0ex}[\value{CEht}\Ctenthex][\Cscr]{}\protect\end{array%
}\protect\right.\protect\label{eq:Paradox-Paradox1}\protect\end{%
eqnarray}\setcounter{CEht}{10}Something has gone wrong! Instead 
of ending up with the star at rest, we find that it is now ``dri%
fting''\mbox{$\!$}. What has happened?\par One can repeat and ch%
eck the algebraic calculations above as many times and in as man%
y ways as one wishes; but the result (\protect\ref{eq:Paradox-Pa%
radox1}) is not a computational error. We can check its self-con%
sistency by noting that, for any four-momentum of a particle of 
mass~$m$, the identity \mbox{$\protect\displaystyle p^{\:\!\mu}p%
_\mu\equiv{E}{}^{\raisebox{-0.25ex}{$\scriptstyle2\!$}}-{{%
\protect\mbox{\protect\boldmath{$p$}}}}{}^{\raisebox{-0.25ex}{$%
\scriptstyle2$}}={m}{}^{\raisebox{-0.25ex}{$\scriptstyle2$}}$} s%
hould be satisfied---as it indeed is for the components listed i%
n (\protect\ref{eq:Paradox-Paradox1}). Moreover, if one simply c%
hanges the order of the final two boosts in (\protect\ref{eq:Par%
adox-Paradox1}), then one finds \setcounter{Ceqindent}{0}\protect
\begin{eqnarray}\hspace{-1.3ex}&\displaystyle B_x(-v_0)^{\,}B_y(%
-v_0)^{\,}B_y(v_0)^{\,}B_x(v_0)\!\left(\begin{array}{c}m\\0\\0\\%
0\end{array}\right)=\left(\begin{array}{c}m\\0\\0\\0\end{array}%
\right)\mbox{$\!$},\protect\nonumber\setlength{\Cscr}{\value{CEh%
t}\Ctenthex}\addtolength{\Cscr}{-1.0ex}\protect\raisebox{0ex}[%
\value{CEht}\Ctenthex][\Cscr]{}\protect\end{eqnarray}\setcounter
{CEht}{10}which would be unlikely to be true had we made any tri%
vial error in computing any of the boost matrices.\par Let us th%
erefore try to find out where our intuition has led us astray in 
the calculation (\protect\ref{eq:Paradox-Paradox1}), by breaking 
it down into smaller parts. We already know what happens to the 
components of the four-momentum of a particle, originally at res%
t, when we subject our system of co\-ordinates to a single Loren%
tz boost, so let us consider instead the effect of the first 
\mbox{}\protect\/{\protect\em two\protect\/} boosts in (\protect
\ref{eq:Paradox-Paradox1}), namely, $B_x(v_0)$ followed by $B_y(%
v_0)$. If we stop our calculation at this point, we find 
\setcounter{Ceqindent}{0}\protect\begin{eqnarray}\protect\left.%
\protect\begin{array}{rcl}\protect\displaystyle\hspace{-1.3ex}&%
\protect\displaystyle B_y(v_0)^{\,}B_x(v_0)\!\left(\begin{array}%
{c}m\\0\\0\\0\end{array}\right)=\left(\begin{array}{c}m\gamma_0^%
2\\\!\!-m\gamma_0^{\protect\mbox{}}v_0\\\!\!-m\gamma_0^2v_0\\0%
\end{array}\right)\mbox{$\!$}.\setlength{\Cscr}{\value{CEht}%
\Ctenthex}\addtolength{\Cscr}{-1.0ex}\protect\raisebox{0ex}[%
\value{CEht}\Ctenthex][\Cscr]{}\protect\end{array}\protect\right
.\protect\label{eq:Paradox-Paradox2}\protect\end{eqnarray}%
\setcounter{CEht}{10}Now, since we have boosted the \mbox{}%
\protect\/{\protect\em Enterprise\protect\/}\ in the positive-$x%
$ and positive-$y$ directions, we expect that the star will be m%
oving (relative to the \mbox{}\protect\/{\protect\em Enterprise%
\protect\/}) with a negative velocity in the $x$- and $y$-direct%
ions; and this is borne out by the result (\protect\ref{eq:Parad%
ox-Paradox2}). However, we are surprised to find that the $x$- a%
nd $y$-velocities \mbox{}\protect\/{\protect\em are not equal%
\protect\/}, despite us boosting the \mbox{}\protect\/{\protect
\em Enterprise\protect\/}\ by the same velocity $v_0$ in each di%
rection! Indeed, making use of the relation (\protect\ref{eq:Pre%
lim-VfromP}) with the components (\protect\ref{eq:Paradox-Parado%
x2}) of \mbox{$\protect\displaystyle p^{\:\!\mu}\!$}, we find th%
at the components of the three-velocity of the star, relative to 
the \mbox{}\protect\/{\protect\em Enterprise\protect\/}, are giv%
en by \setcounter{Ceqindent}{0}\protect\begin{eqnarray}\protect
\left.\protect\begin{array}{rcl}\protect\displaystyle\hspace{-1.%
3ex}&\protect\displaystyle v_x=-\mbox{$\protect\displaystyle
\protect\frac{v_0}{\gamma_0}$},\hspace{12ex}v_y=-v_0.\setlength{%
\Cscr}{\value{CEht}\Ctenthex}\addtolength{\Cscr}{-1.0ex}\protect
\raisebox{0ex}[\value{CEht}\Ctenthex][\Cscr]{}\protect\end{array%
}\protect\right.\protect\label{eq:Paradox-VParadox2}\protect\end
{eqnarray}\setcounter{CEht}{10}Thus, the second ($y$) boost has 
been fully effective---but it has, in the process, reduced the v%
elocity of the first ($x$) boost.\par Let us put this unexpected 
asymmetry to one side, for the moment, and return to our first p%
erplexing result, namely, the nonzero velocity represented by 
\mbox{Eq.~\mbox{$\protect\displaystyle\!$}}(\protect\ref{eq:Para%
dox-Paradox1}). We have found that the application of $B_x(v_0)$ 
and then $B_y(v_0)$ to the \mbox{}\protect\/{\protect\em Enterpr%
ise\protect\/}\ leads to the star having the velocity components 
(\protect\ref{eq:Paradox-VParadox2}) (relative to the \mbox{}%
\protect\/{\protect\em Enterprise\protect\/}). Let us now consid%
er the final two boosts in \mbox{Eq.~\mbox{$\protect\displaystyle
\!$}}(\protect\ref{eq:Paradox-Paradox1}), namely, $B_x(-v_0)$ fo%
llowed by $B_y(-v_0)$. Instead of applying them after the first 
two boosts, let us instead apply them to the \mbox{}\protect\/{%
\protect\em original\protect\/} \mbox{}\protect\/{\protect\em En%
terprise\protect\/}, which was at rest relative to the star. The 
effect of these two boosts on the components of the four-momentu%
m of the star, in this modified scenario, would be \setcounter{C%
eqindent}{0}\protect\begin{eqnarray}\protect\left.\protect\begin
{array}{rcl}\protect\displaystyle\hspace{-1.3ex}&\protect
\displaystyle B_y(-v_0)^{\,}B_x(-v_0)\!\left(\begin{array}{c}m\\%
0\\0\\0\end{array}\right)=\left(\begin{array}{c}m\gamma_0^2\\m%
\gamma_0^{\protect\mbox{}}v_0\\m\gamma_0^2v_0\\0\end{array}\right
)\mbox{$\!$},\setlength{\Cscr}{\value{CEht}\Ctenthex}\addtolength
{\Cscr}{-1.0ex}\protect\raisebox{0ex}[\value{CEht}\Ctenthex][%
\Cscr]{}\protect\end{array}\protect\right.\protect\label{eq:Para%
dox-Paradox3}\protect\end{eqnarray}\setcounter{CEht}{10}leading 
to the velocity components \setcounter{Ceqindent}{0}\protect
\begin{eqnarray}\protect\left.\protect\begin{array}{rcl}\protect
\displaystyle\hspace{-1.3ex}&\protect\displaystyle v_x=\mbox{$%
\protect\displaystyle\protect\frac{v_0}{\gamma_0}$},\hspace{12ex%
}v_y=v_0.\setlength{\Cscr}{\value{CEht}\Ctenthex}\addtolength{%
\Cscr}{-1.0ex}\protect\raisebox{0ex}[\value{CEht}\Ctenthex][\Cscr
]{}\protect\end{array}\protect\right.\protect\label{eq:Paradox-V%
Paradox3}\protect\end{eqnarray}\setcounter{CEht}{10}Thus, compar%
ing (\protect\ref{eq:Paradox-VParadox2}) and (\protect\ref{eq:Pa%
radox-VParadox3}), we find that applying $B_x(-v_0)$ and then $B%
_y(-v_0)$ results in the exact opposite velocity to that obtaine%
d by applying $B_x(v_0)$ and then $B_y(v_0)$. (We would, of cour%
se, expect that this \mbox{}\protect\/{\protect\em would\protect
\/} be the case---but, given the problems we are having, it is e%
ssential to ensure that we do not make intuitive assumptions wit%
hout testing them mathematically.)\par We now find that our orig%
inal result, \mbox{Eq.~\mbox{$\protect\displaystyle\!$}}(\protect
\ref{eq:Paradox-Paradox1}), has not been clarified in the least. 
 For we have shown that our sequence of four boosts can be broke%
n down into a boost by the velocity components (\protect\ref{eq:%
Paradox-VParadox2}) (let us, for definiteness, refer to this thr%
ee-velocity as ${\protect\mbox{\protect\boldmath{$v$}}}_{xy}$), 
followed by a boost by the velocity components (\protect\ref{eq:%
Paradox-VParadox3}) (namely, $-{\protect\mbox{\protect\boldmath{%
$v$}}}_{xy}$). But surely Einstein's very derivation of the Lore%
ntz transformation guarantees us that a boost by any velocity ${%
\protect\mbox{\protect\boldmath{$v$}}}$, followed by a boost by 
$-{\protect\mbox{\protect\boldmath{$v$}}}$, must return us to th%
e original inertial frame? How, then, can we make any sense of t%
he result (\protect\ref{eq:Paradox-Paradox1}), which seems to im%
ply that \setcounter{Ceqindent}{0}\protect\begin{eqnarray}%
\protect\left.\protect\begin{array}{rcl}\protect\displaystyle
\hspace{-1.3ex}&\protect\displaystyle B(-{\protect\mbox{\protect
\boldmath{$v$}}}_{xy})^{\,}B({\protect\mbox{\protect\boldmath{$v%
$}}}_{xy})\neq I?\setlength{\Cscr}{\value{CEht}\Ctenthex}%
\addtolength{\Cscr}{-1.0ex}\protect\raisebox{0ex}[\value{CEht}%
\Ctenthex][\Cscr]{}\protect\end{array}\protect\right.\protect
\label{eq:Paradox-PlusMinus}\protect\end{eqnarray}\setcounter{CE%
ht}{10}\par Let us, again, put this problem to one side, and ins%
tead try the following tack: What if we were to perform four boo%
sts, again in the $+x$, $+y$, $-x$, and $-y$ directions respecti%
vely, but now adjusting the magnitude of each boost velocity so 
as to maintain some sort of control over the resultant overall v%
elocity? Let us again start with a boost by velocity $v_0$ in th%
e positive-$x$ direction: \setcounter{Ceqindent}{0}\protect\begin
{eqnarray}\hspace{-1.3ex}&\displaystyle B_x(v_0)\!\left(\begin{a%
rray}{c}m\\0\\0\\0\end{array}\right)=\left(\begin{array}{c}m%
\gamma_0\\\!\!-m\gamma_0v_0\\0\\0\end{array}\right)\mbox{$\!$}.%
\protect\nonumber\setlength{\Cscr}{\value{CEht}\Ctenthex}%
\addtolength{\Cscr}{-1.0ex}\protect\raisebox{0ex}[\value{CEht}%
\Ctenthex][\Cscr]{}\protect\end{eqnarray}\setcounter{CEht}{10}We 
now apply a boost by some velocity $v_1$ (not equal to $v_0$) in 
the positive-$y$ direction: \setcounter{Ceqindent}{0}\protect
\begin{eqnarray}\hspace{-1.3ex}&\displaystyle B_y(v_1)^{\,}B_x(v%
_0)\!\left(\begin{array}{c}m\\0\\0\\0\end{array}\right)=\left(%
\begin{array}{c}m\gamma_0\gamma_1\\\!\!-m\gamma_0v_0\\\!\!-m%
\gamma_0\gamma_1v_1\\0\end{array}\right)\mbox{$\!$}.\protect
\nonumber\setlength{\Cscr}{\value{CEht}\Ctenthex}\addtolength{%
\Cscr}{-1.0ex}\protect\raisebox{0ex}[\value{CEht}\Ctenthex][\Cscr
]{}\protect\end{eqnarray}\setcounter{CEht}{10}Let us now adjust 
\mbox{$\protect\displaystyle v_{1\!}$} so that the overall veloc%
ity has equal components in the $x$- and $y$-directions (as was 
our original intention). We can do this by ensuring that \mbox{$%
\protect\displaystyle p^{\:\!x\!}$} and \mbox{$\protect
\displaystyle p^{\:\!y\!}$} are equal---namely, by insisting tha%
t \setcounter{Ceqindent}{0}\protect\begin{eqnarray}\hspace{-1.3e%
x}&\displaystyle\gamma_1v_1=v_0.\protect\nonumber\setlength{\Cscr
}{\value{CEht}\Ctenthex}\addtolength{\Cscr}{-1.0ex}\protect
\raisebox{0ex}[\value{CEht}\Ctenthex][\Cscr]{}\protect\end{eqnar%
ray}\setcounter{CEht}{10}After some algebra, one finds that this 
is satisfied for \setcounter{Ceqindent}{0}\protect\begin{eqnarra%
y}\protect\left.\protect\begin{array}{rcl}\protect\displaystyle
\hspace{-1.3ex}&\protect\displaystyle v_1=\mbox{$\protect
\displaystyle\protect\frac{\gamma_0v_0}{\sqrt{\raisebox{-0.15ex}%
[1.5ex][0ex]{$2{\gamma}{}^{ \raisebox{-0.25ex}{$\scriptstyle2$}}%
_0-1$}}}$},\hspace{12ex}\gamma_1=\mbox{$\protect\displaystyle
\protect\frac{\sqrt{\raisebox{-0.15ex}[1.5ex][0ex]{$2{\gamma}{}^%
{ \raisebox{-0.25ex}{$\scriptstyle2$}}_0-1$}}}{\gamma_0}$},%
\setlength{\Cscr}{\value{CEht}\Ctenthex}\addtolength{\Cscr}{-1.0%
ex}\protect\raisebox{0ex}[\value{CEht}\Ctenthex][\Cscr]{}\protect
\end{array}\protect\right.\protect\label{eq:Paradox-V1Solution}%
\protect\end{eqnarray}\setcounter{CEht}{10}so that \setcounter{C%
eqindent}{0}\protect\begin{eqnarray}\hspace{-1.3ex}&\displaystyle
B_y(v_1)^{\,}B_x(v_0)\!\left(\begin{array}{c}m\\0\\0\\0\end{arra%
y}\right)=\left(\begin{array}{c}m\sqrt{\raisebox{-0.15ex}[1.5ex]%
[0ex]{$2{\gamma}{}^{ \raisebox{-0.25ex}{$\scriptstyle2$}}_0-1$}}%
\\\!\!-m\gamma_0v_0\\\!\!-m\gamma_0v_0\\0\end{array}\right)\mbox
{$\!$}.\protect\nonumber\setlength{\Cscr}{\value{CEht}\Ctenthex}%
\addtolength{\Cscr}{-1.0ex}\protect\raisebox{0ex}[\value{CEht}%
\Ctenthex][\Cscr]{}\protect\end{eqnarray}\setcounter{CEht}{10}%
\par Let us now apply a boost by velocity $v_2$ in the negative-%
$x$ direction: \setcounter{Ceqindent}{0}\protect\begin{eqnarray}%
\hspace{-1.3ex}&\displaystyle B_x(-v_2)^{\,}B_y(v_1)^{\,}B_x(v_0%
)\!\left(\begin{array}{c}m\\0\\0\\0\end{array}\right)=\left(%
\begin{array}{c}m\gamma_2(\sqrt{\raisebox{-0.15ex}[1.5ex][0ex]{$%
2{\gamma}{}^{ \raisebox{-0.25ex}{$\scriptstyle2$}}_0-1$}}-\gamma
_0v_0v_2)\\\!\!-m\gamma_2(\gamma_0v_0-v_2\sqrt{\raisebox{-0.15ex%
}[1.5ex][0ex]{$2{\gamma}{}^{ \raisebox{-0.25ex}{$\scriptstyle2$}%
}_0-1$}})\\\!\!-m\gamma_0v_0\\0\end{array}\right)\mbox{$\!$}.%
\protect\nonumber\setlength{\Cscr}{\value{CEht}\Ctenthex}%
\addtolength{\Cscr}{-1.0ex}\protect\raisebox{0ex}[\value{CEht}%
\Ctenthex][\Cscr]{}\protect\end{eqnarray}\setcounter{CEht}{10}We 
can use this third boost to reduce the $x$-component of the velo%
city to zero by choosing \setcounter{Ceqindent}{0}\protect\begin
{eqnarray}\hspace{-1.3ex}&\displaystyle v_2=v_1=\mbox{$\protect
\displaystyle\protect\frac{\gamma_0v_0}{\sqrt{\raisebox{-0.15ex}%
[1.5ex][0ex]{$2{\gamma}{}^{ \raisebox{-0.25ex}{$\scriptstyle2$}}%
_0-1$}}}$},\protect\nonumber\setlength{\Cscr}{\value{CEht}%
\Ctenthex}\addtolength{\Cscr}{-1.0ex}\protect\raisebox{0ex}[%
\value{CEht}\Ctenthex][\Cscr]{}\protect\end{eqnarray}\setcounter
{CEht}{10}yielding \setcounter{Ceqindent}{0}\protect\begin{eqnar%
ray}\hspace{-1.3ex}&\displaystyle B_x(-v_1)^{\,}B_y(v_1)^{\,}B_x%
(v_0)\!\left(\begin{array}{c}m\\0\\0\\0\end{array}\right)=\left(%
\begin{array}{c}m\gamma_0\\0\\\!\!-m\gamma_0v_0\\0\end{array}%
\right)\mbox{$\!$}.\protect\nonumber\setlength{\Cscr}{\value{CEh%
t}\Ctenthex}\addtolength{\Cscr}{-1.0ex}\protect\raisebox{0ex}[%
\value{CEht}\Ctenthex][\Cscr]{}\protect\end{eqnarray}\setcounter
{CEht}{10}Finally, it is evident that we can apply the fourth bo%
ost by the original velocity $v_0$ in the negative-$y$ direction%
, resulting in \setcounter{Ceqindent}{0}\protect\begin{eqnarray}%
\hspace{-1.3ex}&\displaystyle B_y(-v_0)^{\,}B_x(-v_1)^{\,}B_y(v_%
1)^{\,}B_x(v_0)\!\left(\begin{array}{c}m\\0\\0\\0\end{array}%
\right)=\left(\begin{array}{c}m\\0\\0\\0\end{array}\right)\mbox{%
$\!$}.\protect\nonumber\setlength{\Cscr}{\value{CEht}\Ctenthex}%
\addtolength{\Cscr}{-1.0ex}\protect\raisebox{0ex}[\value{CEht}%
\Ctenthex][\Cscr]{}\protect\end{eqnarray}\setcounter{CEht}{10}We 
finally seem to have found a sequence of four boosts, in the $+x%
$, $+y$, $-x$, and $-y$ directions respectively, that returns th%
e \mbox{}\protect\/{\protect\em Enterprise\protect\/}\ to a stat%
e of rest relative to the star at the end of the maneuver: namel%
y, the sequence of boosts \setcounter{Ceqindent}{0}\protect\begin
{eqnarray}\protect\left.\protect\begin{array}{rcl}\protect
\displaystyle\hspace{-1.3ex}&\protect\displaystyle B_y(-v_0)^{\,%
}B_x(-v_1)^{\,}B_y(v_1)^{\,}B_x(v_0)\setlength{\Cscr}{\value{CEh%
t}\Ctenthex}\addtolength{\Cscr}{-1.0ex}\protect\raisebox{0ex}[%
\value{CEht}\Ctenthex][\Cscr]{}\protect\end{array}\protect\right
.\protect\label{eq:Paradox-BoostRestStar}\protect\end{eqnarray}%
\setcounter{CEht}{10}together with the relation (\protect\ref{eq%
:Paradox-V1Solution}) between \mbox{$\protect\displaystyle v_{1%
\!}$} and $v_0$.\par Let us now go back in time to our original 
\mbox{}\protect\/{\protect\em Enterprise\protect\/}, at rest rel%
ative to the nearby star. The crew of the \mbox{}\protect\/{%
\protect\em Enterprise\protect\/}\ had noted that there was a sh%
uttlecraft, of mass $m_s$, moving with velocity $v_s$ in the pos%
itive-$x$ direction; in other words, the components of its four-%
momentum, relative to the \mbox{}\protect\/{\protect\em Enterpri%
se\protect\/}, were \setcounter{Ceqindent}{0}\protect\begin{eqna%
rray}\protect\left.\protect\begin{array}{rcl}\protect
\displaystyle\hspace{-1.3ex}&\protect\displaystyle p^{\:\!\mu\!}%
=\left(\begin{array}{c}m_{s\:\!\!}\gamma_s\\m_{s\:\!\!}\gamma_sv%
_s\\0\\0\end{array}\right)\mbox{$\!$}.\setlength{\Cscr}{\value{C%
Eht}\Ctenthex}\addtolength{\Cscr}{-1.0ex}\protect\raisebox{0ex}[%
\value{CEht}\Ctenthex][\Cscr]{}\protect\end{array}\protect\right
.\protect\label{eq:Paradox-Shuttle}\protect\end{eqnarray}%
\setcounter{CEht}{10}What happens to the components of the four-%
momentum of the shuttlecraft after the sequence of boosts (%
\protect\ref{eq:Paradox-BoostRestStar})? We would \mbox{}\protect
\/{\protect\em expect\protect\/} that they---like those of the s%
tar---would be unchanged. However, if we perform the calculation%
s, we find that \setcounter{Ceqindent}{0}\protect\begin{eqnarray%
}\protect\left.\protect\begin{array}{rcl}\protect\displaystyle
\hspace{-1.3ex}&\protect\displaystyle B_y(-v_0)^{\,}B_x(-v_1)^{%
\,}B_y(v_1)^{\,}B_x(v_0)\!\left(\begin{array}{c}m_{s\:\!\!}\gamma
_s\\m_{s\:\!\!}\gamma_sv_s\\0\\0\end{array}\right)=\left(\begin{%
array}{c}m_{s\:\!\!}\gamma_s\\m_{s\:\!\!}\gamma_sv_s\sqrt{%
\raisebox{-0.15ex}[1.5ex][0ex]{$2{\gamma}{}^{ \raisebox{-0.25ex}%
{$\scriptstyle2$}}_0-1$}}/\gamma_0^2\\m_{s\:\!\!}\gamma_sv_s(1-1%
/\gamma_0^2)\\0\end{array}\right)\mbox{$\!$}.\setlength{\Cscr}{%
\value{CEht}\Ctenthex}\addtolength{\Cscr}{-1.0ex}\protect
\raisebox{0ex}[\value{CEht}\Ctenthex][\Cscr]{}\protect\end{array%
}\protect\right.\protect\label{eq:Paradox-ShuttleRotate}\protect
\end{eqnarray}\setcounter{CEht}{10}We can't seem to take a trick%
! Even though the sequence of boosts (\protect\ref{eq:Paradox-Bo%
ostRestStar}) has left the velocity of the star unchanged, relat%
ive to the \mbox{}\protect\/{\protect\em Enterprise\protect\/}, 
it has \mbox{}\protect\/{\protect\em changed\protect\/} the velo%
city of the shuttlecraft.\par But how can this be possible?\par
\refstepcounter{section}\vspace{1.5\baselineskip}\par{\centering
\bf\thesection. Solutions to the elementary ``paradoxes''\\*[0.5%
\baselineskip]}\protect\indent\label{sect:Solution}Let us now di%
scover the fallacies contained in the ``paradoxes'' described ab%
ove. We shall begin by unraveling our chain of arguments, starti%
ng with the final ``paradox''\mbox{$\!$}, and working our way ba%
ck to the first. By this stage, the reasons for each ``paradox'' 
will be clear. We shall complete this section by listing general 
expressions for the Thomas rotation in arbitrary cases.\par We b%
egin with the result (\protect\ref{eq:Paradox-ShuttleRotate}) fo%
r the final four-momentum of the shuttlecraft. We were surprised 
to find that it differed from the four-momentum (\protect\ref{eq%
:Paradox-Shuttle}) of the shuttlecraft prior to the sequence of 
boosts. However, on closer inspection, we find that the result i%
s not total chaos. In particular, the \mbox{}\protect\/{\protect
\em energy\protect\/} of the shuttlecraft has not changed. This, 
in turn, implies that its \mbox{}\protect\/{\protect\em speed%
\protect\/} is also unchanged---in other words, the final veloci%
ty has the same magnitude as the original velocity, but \mbox{}%
\protect\/{\protect\em it has been rotated in space\protect\/}. 
We can confirm this by using Pythagoras's theorem to find the re%
sultant of the components \mbox{$\protect\displaystyle v_{x\!}$} 
and \mbox{$\protect\displaystyle v_{y\!}$} in (\protect\ref{eq:P%
aradox-ShuttleRotate}); and we indeed find that its magnitude is 
simply $v_s$, the original speed of the shuttlecraft.\par This r%
otation of the spatial axes is what we know as the \mbox{}%
\protect\/{\protect\em Thomas rotation\protect\/}. It almost alw%
ays occurs when we apply a sequence of non-collinear boosts that 
returns us to an inertial frame that is at rest relative to the 
original frame. This rotation had no net effect on the four-mome%
ntum of the star, because the star was at rest---the ``spatial'' 
components of its four-momentum vanished; in contrast, the motio%
n of the shuttlecraft defines a direction in space (the $x$-dire%
ction, in the original inertial frame), that was subject to the 
rotation.\par We can obtain a clearer view of this rotation if w%
e compute not the components of the four-momentum (\protect\ref{%
eq:Paradox-ShuttleRotate}), but rather the entire matrix (%
\protect\ref{eq:Paradox-BoostRestStar}) that is applicable to 
\mbox{}\protect\/{\protect\em arbitrary\protect\/} four-vectors 
in the original frame: \setcounter{Ceqindent}{0}\protect\begin{e%
qnarray}\protect\left.\protect\begin{array}{rcl}\protect
\displaystyle\hspace{-1.3ex}&\protect\displaystyle B_y(-v_0)^{\,%
}B_x(-v_1)^{\,}B_y(v_1)^{\,}B_x(v_0)=\left(\begin{array}{cccc}\,%
\,1&0\raisebox{0ex}[3ex][2ex]{}&0&0\,\,\\\,\,0&\mbox{$\protect
\displaystyle\protect\frac{\sqrt{\raisebox{-0.15ex}[1.5ex][0ex]{%
$2\gamma_0^2-1$}}}{\gamma_0^2}$}\raisebox{0ex}[0ex][2.5ex]{}&\!%
\!-\mbox{$\protect\displaystyle\protect\frac{\gamma_0^2-1}{\gamma
_0^2}$}&0\,\,\\\,\,0&\mbox{$\protect\displaystyle\protect\frac{%
\gamma_0^2-1}{\gamma_0^2}$}&\mbox{$\protect\displaystyle\protect
\frac{\sqrt{\raisebox{-0.15ex}[1.5ex][0ex]{$2\gamma_0^2-1$}}}{%
\gamma_0^2}$}&0\,\,\\\,\,0&0\raisebox{0ex}[3.5ex][1ex]{}&0&1\,\,%
\end{array}\right)\mbox{$\!$}.\setlength{\Cscr}{\value{CEht}%
\Ctenthex}\addtolength{\Cscr}{-1.0ex}\protect\raisebox{0ex}[%
\value{CEht}\Ctenthex][\Cscr]{}\protect\end{array}\protect\right
.\protect\label{eq:Solution-FullMatrix}\protect\end{eqnarray}%
\setcounter{CEht}{10}The \mbox{$\protect\displaystyle2\times2$} 
matrix in the middle of this result is an orthogonal transformat%
ion, resulting in a rotation of the axes of the $x$--$y$ plane b%
y an angle \setcounter{Ceqindent}{0}\protect\begin{eqnarray}%
\protect\left.\protect\begin{array}{rcl}\protect\displaystyle
\hspace{-1.3ex}&\protect\displaystyle\theta=-\mbox{arctan}\!%
\protect\left(\mbox{$\protect\displaystyle\protect\frac{\,\gamma
_0^2-1}{\!\sqrt{\raisebox{-0.15ex}[1.5ex][0ex]{$2\gamma_0^2-1$}}%
}$}\protect\right)\!.\setlength{\Cscr}{\value{CEht}\Ctenthex}%
\addtolength{\Cscr}{-1.0ex}\protect\raisebox{0ex}[\value{CEht}%
\Ctenthex][\Cscr]{}\protect\end{array}\protect\right.\protect
\label{eq:Solution-Angle}\protect\end{eqnarray}\setcounter{CEht}%
{10}In the nonrelativistic limit, the magnitude of $\theta$ appr%
oaches \mbox{$\protect\displaystyle{v}{}^{\raisebox{-0.25ex}{$%
\scriptstyle2\!$}}$} radians (i.e., \mbox{$\protect\displaystyle
{v}{}^{\raisebox{-0.25ex}{$\scriptstyle2$}}\!/{c}{}^{\raisebox{-%
0.25ex}{$\scriptstyle2\!$}}$} in conventional units), and so is 
completely negligible for terrestrial applications. (Even if 
\mbox{$\protect\displaystyle v_{0\!}$} is set to the Earth's orb%
ital velocity around the Sun, the Thomas rotation angle amounts 
to a mere 0.004~seconds of arc.) In the ultrarelativistic limit, 
on the other hand, $\theta$ approaches $-90\mbox{$^\circ$}$ for 
this particular sequence of boosts.\par Defining the \mbox{}%
\protect\/{\protect\em direction\protect\/} of the Thomas rotati%
on, however, requires some care. Let us consider the above seque%
nce of boosts from the point of view of an inertial observer, je%
ttisoned from the \mbox{}\protect\/{\protect\em Enterprise%
\protect\/}\ before the sequence of boosts commenced, who \mbox{%
}\protect\/{\protect\em remained\protect\/} at rest relative to 
the star (and the distant ``fixed stars'') throughout the proced%
ure. Relative to this fixed observer, the \mbox{}\protect\/{%
\protect\em Enterprise\protect\/}'s velocity rotated in the dire%
ction \mbox{$\protect\displaystyle+x\rightarrow+y$}. The velocit%
y of the shuttlecraft, \mbox{}\protect\/{\protect\em as seen by 
the\protect\/} \mbox{}\protect\/{\protect\em Enterprise\protect
\/}, was rotated in this same direction. This means that, relati%
ve to the fixed observer, the axes of the \mbox{}\protect\/{%
\protect\em Enterprise\protect\/}'s co\-ordinate system rotated 
in the \mbox{}\protect\/{\protect\em opposite\protect\/} directi%
on to its orbital rotation, as indicated by the minus sign in 
\mbox{Eq.~\mbox{$\protect\displaystyle\!$}}(\protect\ref{eq:Solu%
tion-Angle}). This is a general feature of the Thomas rotation.%
\par Nonrelativistic physics has conditioned us to assume that C%
artesian co\-ordinate systems can be defined in space, in such a 
way that all inertial observers ``agree'' on the directions of t%
he axes. The Thomas rotation demonstrates that this assumption r%
equires an operational definition, as Einstein showed was necess%
ary to clarify our understanding of the physics of relativity. F%
or example, say that observer $A$ defines a set of Cartesian axe%
s. If observer $B$ is at rest relative to~$A$, then $B$ can alig%
n her axes to ``agree'' with those of~$A$. If observer $A$ remai%
ns at rest, but observer $B$ is boosted to some finite velocity 
relative to $A$, by one boost or by a sequence of boosts, then 
\mbox{}\protect\/{\protect\em the resultant orientation of $B$'s 
axes depends on the particular sequence of boosts used\protect\/%
}. If such boosts are at all times in the same direction (relati%
ve to $A$, say), then it is meaningful to say that $B$'s axes ar%
e still aligned with $A$'s, in the sense that if we apply any se%
quence of boosts to $B$ that is at all times collinear with this 
direction, that returns $B$ to rest with respect to $A$, then th%
eir axes will be found to still point in the same directions. On 
the other hand, if $B$ is, at any two times, subject to boosts i%
n \mbox{}\protect\/{\protect\em different\protect\/} directions, 
then a sequence of boosts bringing $B$ back to rest relative to 
$A$ will, in general, lead to $B$ finding her axes rotated relat%
ive to $A$'s (unless the sequence of boosts ``backtracked'' prec%
isely the original sequence).\par To discuss the general case we 
need the expression for a (simple) boost by velocity $|{\protect
\mbox{\protect\boldmath{$v$}}}|$ in the direction of~${\protect
\mbox{\protect\boldmath{$v$}}}$. We can obtain it most simply by 
noting that our original boost transformation along the $x$-axis%
, $B_x(v)$ of \mbox{Eq.~\mbox{$\protect\displaystyle\!$}}(%
\protect\ref{eq:Prelim-Bx}), is an archetypical simple boost. If 
we rewrite $B_x(v)$ in three-covariant notation (i.e., in terms 
of three-vectors and three-vector operations, rather than indivi%
dual components), then we know from vector analysis that the res%
ult will be the boost we require for ${\protect\mbox{\protect
\boldmath{$v$}}}$ in an arbitrary direction. It is now straightf%
orward to verify that$^{\ref{cit:Jackson1999}}$ \setcounter{Ceqi%
ndent}{0}\protect\begin{eqnarray}\protect\left.\protect\begin{ar%
ray}{rcl}\protect\displaystyle\hspace{-1.3ex}&\protect
\displaystyle t'=\gamma t-\gamma({\protect\mbox{\protect\boldmath
{$v$}}}\mbox{$\hspace{0.2ex}\cdot\hspace{0.2ex}$}{\protect\mbox{%
\protect\boldmath{$x$}}}),\hspace{12ex}{\protect\mbox{\protect
\boldmath{$x$}}}'={\protect\mbox{\protect\boldmath{$x$}}}-\gamma
t{\protect\mbox{\protect\boldmath{$v$}}}+\mbox{$\protect
\displaystyle\protect\frac{{\gamma}{}^{\raisebox{-0.25ex}{$%
\scriptstyle2$}}}{\gamma+1}$}({\protect\mbox{\protect\boldmath{$%
v$}}}\mbox{$\hspace{0.2ex}\cdot\hspace{0.2ex}$}{\protect\mbox{%
\protect\boldmath{$x$}}}){\protect\mbox{\protect\boldmath{$v$}}}%
,\setlength{\Cscr}{\value{CEht}\Ctenthex}\addtolength{\Cscr}{-1.%
0ex}\protect\raisebox{0ex}[\value{CEht}\Ctenthex][\Cscr]{}%
\protect\end{array}\protect\right.\protect\label{eq:Solution-Boo%
stCov}\protect\end{eqnarray}\setcounter{CEht}{10}is equivalent t%
o (\protect\ref{eq:Prelim-Bx}) for \mbox{$\protect\displaystyle{%
\protect\mbox{\protect\boldmath{$v$}}}=(v,0,0)$}; and thus (%
\protect\ref{eq:Solution-BoostCov}) is the simple boost operatio%
n $B({\protect\mbox{\protect\boldmath{$v$}}})$ that we are seeki%
ng. (It is a straightforward calculation to confirm that the com%
ponent of ${\protect\mbox{\protect\boldmath{$x$}}}$ in the direc%
tion of ${\protect\mbox{\protect\boldmath{$v$}}}$ satisfies the 
usual Lorentz transformation: \setcounter{Ceqindent}{0}\protect
\begin{eqnarray}\hspace{-1.3ex}&\displaystyle\mbox{$\protect
\displaystyle\protect\frac{{\protect\mbox{\protect\boldmath{$v$}%
}}\mbox{$\hspace{0.2ex}\cdot\hspace{0.2ex}$}{\protect\mbox{%
\protect\boldmath{$x$}}}'}{v}$}=\mbox{$\protect\displaystyle
\protect\frac{\gamma{\protect\mbox{\protect\boldmath{$v$}}}\mbox
{$\hspace{0.2ex}\cdot\hspace{0.2ex}$}{\protect\mbox{\protect
\boldmath{$x$}}}}{v}$}-\gamma vt,\protect\nonumber\setlength{%
\Cscr}{\value{CEht}\Ctenthex}\addtolength{\Cscr}{-1.0ex}\protect
\raisebox{0ex}[\value{CEht}\Ctenthex][\Cscr]{}\protect\end{eqnar%
ray}\setcounter{CEht}{10}and that the components of ${\protect
\mbox{\protect\boldmath{$x$}}}$ normal to the velocity ${\protect
\mbox{\protect\boldmath{$v$}}}$ are unchanged: \setcounter{Ceqin%
dent}{0}\protect\begin{eqnarray}\hspace{-1.3ex}&\displaystyle{%
\protect\mbox{\protect\boldmath{$v$}}}\mbox{$\times$}{\protect
\mbox{\protect\boldmath{$x$}}}^{\prime\!}={\protect\mbox{\protect
\boldmath{$v$}}}\mbox{$\times$}{\protect\mbox{\protect\boldmath{%
$x$}}},\protect\nonumber\setlength{\Cscr}{\value{CEht}\Ctenthex}%
\addtolength{\Cscr}{-1.0ex}\protect\raisebox{0ex}[\value{CEht}%
\Ctenthex][\Cscr]{}\protect\end{eqnarray}\setcounter{CEht}{10}si%
nce the component of ${\protect\mbox{\protect\boldmath{$x$}}}$ o%
r \mbox{$\protect\displaystyle{\protect\mbox{\protect\boldmath{$%
x$}}}^{\prime\!}$} in the direction of ${\protect\mbox{\protect
\boldmath{$v$}}}$ does not contribute to the cross product.) Wri%
tten out in matrix form, we have \setcounter{Ceqindent}{0}%
\protect\begin{eqnarray}\protect\left.\protect\begin{array}{rcl}%
\protect\displaystyle\hspace{-1.3ex}&\protect\displaystyle B({%
\protect\mbox{\protect\boldmath{$v$}}})=\left(\begin{array}{cccc%
}\raisebox{0ex}[3ex][3ex]{}\gamma&\!\!-\gamma v_x&\!\!-\gamma v_%
y&\!\!-\gamma v_z\\\!\!-\gamma v_x\raisebox{0ex}[3ex][3ex]{}&1+%
\mbox{$\protect\displaystyle\protect\frac{\gamma^2v_x^2}{\gamma+%
1}$}&\mbox{$\protect\displaystyle\protect\frac{\gamma^2v_xv_y}{%
\gamma+1}$}&\mbox{$\protect\displaystyle\protect\frac{\gamma^2v_%
xv_z}{\gamma+1}$}\\\!\!-\gamma v_y\raisebox{0ex}[3ex][3ex]{}&%
\mbox{$\protect\displaystyle\protect\frac{\gamma^2v_xv_y}{\gamma
+1}$}&1+\mbox{$\protect\displaystyle\protect\frac{\gamma^2v_y^2}%
{\gamma+1}$}&\mbox{$\protect\displaystyle\protect\frac{\gamma^2v%
_yv_z}{\gamma+1}$}\\\!\!-\gamma v_z\raisebox{0ex}[3ex][3ex]{}&%
\mbox{$\protect\displaystyle\protect\frac{\gamma^2v_xv_z}{\gamma
+1}$}&\mbox{$\protect\displaystyle\protect\frac{\gamma^2v_yv_z}{%
\gamma+1}$}&1+\mbox{$\protect\displaystyle\protect\frac{\gamma^2%
v_z^2}{\gamma+1}$}\end{array}\right)\mbox{$\!$}.\setlength{\Cscr
}{\value{CEht}\Ctenthex}\addtolength{\Cscr}{-1.0ex}\protect
\raisebox{0ex}[\value{CEht}\Ctenthex][\Cscr]{}\protect\end{array%
}\protect\right.\protect\label{eq:Solution-BV}\protect\end{eqnar%
ray}\setcounter{CEht}{10}A sequence of two such boosts, which ar%
e in different directions, is not a simple boost, but is rather 
a combination of a rotation and a simple boost. If we consider $%
B({\protect\mbox{\protect\boldmath{$v$}}}_1)$ followed by $B({%
\protect\mbox{\protect\boldmath{$v$}}}_2)$, and denote the veloc%
ity of the boost implied by the composite transformation as ${%
\protect\mbox{\protect\boldmath{$v$}}}_{12}$, then the mathemati%
cal expression of this observation is that \setcounter{Ceqindent%
}{0}\protect\begin{eqnarray}\hspace{-1.3ex}&\displaystyle B({%
\protect\mbox{\protect\boldmath{$v$}}}_2)^{\,}B({\protect\mbox{%
\protect\boldmath{$v$}}}_1)=B({\protect\mbox{\protect\boldmath{$%
v$}}}_{12})^{\,}R({\protect\mbox{\protect\boldmath{$v$}}}_1,{%
\protect\mbox{\protect\boldmath{$v$}}}_2),\protect\nonumber
\setlength{\Cscr}{\value{CEht}\Ctenthex}\addtolength{\Cscr}{-1.0%
ex}\protect\raisebox{0ex}[\value{CEht}\Ctenthex][\Cscr]{}\protect
\end{eqnarray}\setcounter{CEht}{10}where $R({\protect\mbox{%
\protect\boldmath{$v$}}}_1,{\protect\mbox{\protect\boldmath{$v$}%
}}_2)$ is a spatial rotation, depending on the velocities \mbox{%
$\protect\displaystyle{\protect\mbox{\protect\boldmath{$v$}}}_{1%
\!}$} and \mbox{$\protect\displaystyle{\protect\mbox{\protect
\boldmath{$v$}}}_2$}.\par Let us now return to the ``unexpected 
asymmetry'' in the result (\protect\ref{eq:Paradox-VParadox2}), 
namely, the fact that a boost by the velocity $v_0$ in the $x$-d%
irection, followed by a boost by $v_0$ in the $y$-direction, lea%
ds to \mbox{$\protect\displaystyle v_x\neq v_y$} relative to the 
original frame of reference. We can understand the result \mbox{%
$\protect\displaystyle v_y=-v_0$} by the following argument: Ima%
gine that, after the $x$-boost, there is an object that is obser%
ved to be at rest. Applying the $y$-boost to ourselves, we have 
no choice but to observe this object moving with velocity $-v_0$ 
in the $y$-direction. This same argument must apply to any objec%
t in the original frame which had no velocity in the $y$-directi%
on.\par The reduction in the $x$-velocity by the $y$-boost seems 
counterintuitive, but a little thought makes sense of it. We kno%
w, from Einstein's ingenious arguments, that lengths of rods per%
pendicular to a velocity vector are unchanged by the relative mo%
tion. But lengths are nothing more than differences in positions%
; and positions are themselves the spatial components of the fou%
r-vector $x^\mu$\mbox{$\!$}. Thus, taking into account the unive%
rsality of Lorentz covariance, Einstein's arguments imply that, 
for \mbox{}\protect\/{\protect\em any\protect\/} four-vector, th%
e spatial components perpendicular to the boost velocity are unc%
hanged by the boost, as can be verified from \mbox{Eq.~\mbox{$%
\protect\displaystyle\!$}}(\protect\ref{eq:Solution-BoostCov}). 
But the \mbox{}\protect\/{\protect\em spatial momentum\protect\/%
} components are simply the spatial components of the four-vecto%
r \mbox{$\protect\displaystyle p^{\:\!\mu}$}; therefore, \mbox{$%
\protect\displaystyle{p}{}^{\raisebox{-0.25ex}{$\scriptstyle\:\!%
x\!$}}$} and \mbox{$\protect\displaystyle{p}{}^{\raisebox{-0.25e%
x}{$\scriptstyle\:\!z\!$}}$} must be unaltered by a $y$-boost. A%
nd, indeed, the result (\protect\ref{eq:Paradox-Paradox2}) shows 
us that the $x$-momentum of the star \mbox{}\protect\/{\protect
\em was\protect\/} unchanged by the second boost: it remained 
\mbox{$\protect\displaystyle-m\gamma_0v_0$}. Rather, the \mbox{}%
\protect\/{\protect\em energy\protect\/} of the star increased (%
due to its new $y$-velocity); and hence, by \mbox{Eq.~\mbox{$%
\protect\displaystyle\!$}}(\protect\ref{eq:Prelim-VfromP}), its 
$x$-velocity \mbox{}\protect\/{\protect\em decreased\protect\/}. 
To have an object maintain its momentum, but lose velocity, is n%
onrelativistically counterintuitive; but one can make sense of i%
t by remembering that all velocities must remain smaller than th%
at of light, and so for a large enough boost in the $y$-directio%
n, any original velocity in the $x$-direction must be ``quenched%
'' (although not its momentum!).\par This asymmetry tells us tha%
t the non-commutativity of two non-collinear boosts is more comp%
licated than is widely appreciated: one not only finds a relativ%
e Thomas rotation between the two resulting frames, but furtherm%
ore \mbox{}\protect\/{\protect\em the resulting frames are not e%
ven moving with the same velocity\protect\/}. This is the source 
of the result expressed in \mbox{Eq.~\mbox{$\protect\displaystyle
\!$}}(\protect\ref{eq:Paradox-Paradox1}). There we considered a 
sequence of four boosts which we naively expected to return us t%
o our initial state of motion. However, the first two boosts do 
not combine to give a pure Lorentz boost, but rather involve a T%
homas rotation. This rotation is not compensated by the later bo%
osts---indeed, there is a \mbox{}\protect\/{\protect\em further%
\protect\/} rotation in the same direction. Thus, we should not 
be surprised that the sequence of four boosts gives the counteri%
ntuitive result of \mbox{Eq.~\mbox{$\protect\displaystyle\!$}}(%
\protect\ref{eq:Paradox-Paradox1}).\par This fundamental asymmet%
ry is ``hidden'' in many introductory accounts of the addition o%
f non-collinear velocities, by means of a judicious mixing of an 
\mbox{}\protect\/{\protect\em active\protect\/} transformation f%
or one velocity (i.e., the object is considered to be boosted, w%
ith we as observers being kept at rest) together with a \mbox{}%
\protect\/{\protect\em passive\protect\/} transformation for the 
other (i.e., we as observers are being boosted), rather than two 
successive passive transformations as used in this paper. This 
``trick'' gives the illusion of a greater degree of symmetry tha%
n is generally the case. (Einstein's seminal 1905 paper uses thi%
s ``trick''\mbox{$\!$}.)\par All of these various points must be 
kept in mind if one wishes to analyze Thomas rotations in full g%
enerality. Any ``closed'' sequence of finite boosts (i.e., that 
returns us to a frame at rest relative to the original frame) wi%
ll, in general, result in a Thomas rotation. Any such closed seq%
uence may be broken down into a succession of closed sequences, 
each consisting of three boosts, in the same way that any arbitr%
ary polygon (not necessarily planar) can be broken down into a 
``triangular mesh'' by the addition of internal edges. Thus, the 
basic ``building block'' of a finite Thomas rotation is a sequen%
ce of three pure boosts: the first two are arbitrary, and the th%
ird must be chosen so as to make the sequence ``closed''\mbox{$%
\!$}. The first two velocities, then, determine the Thomas rotat%
ion for this ``building block''\mbox{$\!$}. Complicating such ca%
lculations, however, is the fact that the ``sum'' \mbox{$\protect
\displaystyle{\protect\mbox{\protect\boldmath{$v$}}}_{12}$} of t%
wo arbitrary velocities \mbox{$\protect\displaystyle{\protect
\mbox{\protect\boldmath{$v$}}}_{1^{\!}}$} and \mbox{$\protect
\displaystyle{\protect\mbox{\protect\boldmath{$v$}}}_{2^{\!}}$} 
is, in the general case, quite a complicated function of the fir%
st two velocities: \setcounter{Ceqindent}{0}\protect\begin{eqnar%
ray}\protect\left.\protect\begin{array}{rcl}\protect\displaystyle
\hspace{-1.3ex}&\protect\displaystyle{\protect\mbox{\protect
\boldmath{$v$}}}_{12}=\mbox{$\protect\displaystyle\protect\frac{%
1}{\gamma_2(1+{\protect\mbox{\protect\boldmath{$v$}}}_{1\!}\mbox
{$\hspace{0.2ex}\cdot\hspace{0.2ex}$}{\protect\mbox{\protect
\boldmath{$v$}}}_2)}$}\!\protect\left\{{\protect\mbox{\protect
\boldmath{$v$}}}_1+\gamma_2{\protect\mbox{\protect\boldmath{$v$}%
}}_2+\mbox{$\protect\displaystyle\protect\frac{\gamma_2^2}{\gamma
_{2\!}+1}$}({\protect\mbox{\protect\boldmath{$v$}}}_{1\!}\mbox{$%
\hspace{0.2ex}\cdot\hspace{0.2ex}$}{\protect\mbox{\protect
\boldmath{$v$}}}_2){\protect\mbox{\protect\boldmath{$v$}}}_2%
\protect\right\}\!.\setlength{\Cscr}{\value{CEht}\Ctenthex}%
\addtolength{\Cscr}{-1.0ex}\protect\raisebox{0ex}[\value{CEht}%
\Ctenthex][\Cscr]{}\protect\end{array}\protect\right.\protect
\label{eq:Solution-SumV}\protect\end{eqnarray}\setcounter{CEht}{%
10}(We can clearly see here the asymmetry between the two veloci%
ties \mbox{$\protect\displaystyle{\protect\mbox{\protect\boldmath
{$v$}}}_{1\!}$} and ${\protect\mbox{\protect\boldmath{$v$}}}_2$; 
it is only if \mbox{$\protect\displaystyle{\protect\mbox{\protect
\boldmath{$v$}}}_{1\!}$} and \mbox{$\protect\displaystyle{%
\protect\mbox{\protect\boldmath{$v$}}}_{2\!}$} are collinear tha%
t (\protect\ref{eq:Solution-SumV}) becomes symmetrical under the%
ir interchange, and reproduces the usual formula (\protect\ref{e%
q:Prelim-AddVStd}) for the relativistic addition of velocities, 
as a short calculation shows.) The expression for the Thomas rot%
ation is, in turn, even more complicated. Let us assume that we 
have an arbitrary three-vector ${\protect\mbox{\protect\boldmath
{$r$}}}$ in our initial frame. After the sequence of pure boosts 
$B({\protect\mbox{\protect\boldmath{$v$}}}_1)$, $B({\protect\mbox
{\protect\boldmath{$v$}}}_2)$, and $B(-{\protect\mbox{\protect
\boldmath{$v$}}}_{12})$, the three-vector ${\protect\mbox{%
\protect\boldmath{$r$}}}$ is rotated to \setcounter{CElett}{0}%
\protect\refstepcounter{equation}\protect\label{eq:Solution-Fina%
l}\renewcommand\theequation{\arabic{equation}\alph{CElett}}%
\protect\stepcounter{CElett}\addtocounter{equation}{-1}%
\setcounter{Ceqindent}{0}\protect\begin{eqnarray}\protect\left.%
\protect\begin{array}{rcl}\protect\displaystyle\hspace{-1.3ex}&%
\protect\displaystyle{\protect\mbox{\protect\boldmath{$r$}}}^{%
\prime\!}={\protect\mbox{\protect\boldmath{$r$}}}+\mbox{$\protect
\displaystyle\protect\frac{\gamma_1\gamma_2({\protect\mbox{%
\protect\boldmath{$v$}}}_{1\!}\mbox{$\times$}{\protect\mbox{%
\protect\boldmath{$v$}}}_2)\mbox{$\times$}{\protect\mbox{\protect
\boldmath{$r$}}}-{\protect\mbox{\protect\boldmath{$Q$}}}}{1+%
\gamma_1\gamma_2(1+{\protect\mbox{\protect\boldmath{$v$}}}_{1\!}%
\mbox{$\hspace{0.2ex}\cdot\hspace{0.2ex}$}{\protect\mbox{\protect
\boldmath{$v$}}}_2)}$},\setlength{\Cscr}{\value{CEht}\Ctenthex}%
\addtolength{\Cscr}{-1.0ex}\protect\raisebox{0ex}[\value{CEht}%
\Ctenthex][\Cscr]{}\protect\end{array}\protect\right.\protect
\label{eq:Solution-FinalMain}\protect\end{eqnarray}\setcounter{C%
Eht}{10}where \protect\stepcounter{CElett}\addtocounter{equation%
}{-1}\setcounter{Ceqindent}{0}\protect\begin{eqnarray}\protect
\left.\protect\begin{array}{rcl}\protect\displaystyle\hspace{-1.%
3ex}&\protect\displaystyle{\protect\mbox{\protect\boldmath{$Q$}}%
}\equiv\mbox{$\protect\displaystyle\protect\frac{\gamma_1^2(%
\gamma_2^2-1)({\protect\mbox{\protect\boldmath{$v$}}}_{1\!}\mbox
{$\hspace{0.2ex}\cdot\hspace{0.2ex}$}{\protect\mbox{\protect
\boldmath{$r$}}}){\protect\mbox{\protect\boldmath{$v$}}}_1+\gamma
_2^2(\gamma_1^2-1)({\protect\mbox{\protect\boldmath{$v$}}}_{2\!}%
\mbox{$\hspace{0.2ex}\cdot\hspace{0.2ex}$}{\protect\mbox{\protect
\boldmath{$r$}}}){\protect\mbox{\protect\boldmath{$v$}}}_2-2%
\gamma_1^2\gamma_2^2({\protect\mbox{\protect\boldmath{$v$}}}_{1%
\!}\mbox{$\hspace{0.2ex}\cdot\hspace{0.2ex}$}{\protect\mbox{%
\protect\boldmath{$v$}}}_2)({\protect\mbox{\protect\boldmath{$v$%
}}}_{1\!}\mbox{$\hspace{0.2ex}\cdot\hspace{0.2ex}$}{\protect\mbox
{\protect\boldmath{$r$}}}){\protect\mbox{\protect\boldmath{$v$}}%
}_2}{(\gamma_{1\!}+1)(\gamma_{2\!}+1)}$}.\setlength{\Cscr}{\value
{CEht}\Ctenthex}\addtolength{\Cscr}{-1.0ex}\protect\raisebox{0ex%
}[\value{CEht}\Ctenthex][\Cscr]{}\protect\end{array}\protect
\right.\protect\label{eq:Solution-FinalRemainder}\protect\end{eq%
narray}\setcounter{CEht}{10}\renewcommand\theequation{\arabic{eq%
uation}}(Again, ${\protect\mbox{\protect\boldmath{$Q$}}}$ has no 
particular symmetry under the interchange of \mbox{$\protect
\displaystyle{\protect\mbox{\protect\boldmath{$v$}}}_{1\!}$} and 
${\protect\mbox{\protect\boldmath{$v$}}}_2$.) It can be verified%
, after some algebra, that \mbox{$\protect\displaystyle{\protect
\mbox{\protect\boldmath{$r$}}}^{\prime\:\!2\!}={\protect\mbox{%
\protect\boldmath{$r$}}}^2$}\mbox{$\!$}, i.e., that ${\protect
\mbox{\protect\boldmath{$r$}}}'$ is indeed simply a rotation of 
${\protect\mbox{\protect\boldmath{$r$}}}$ in three-space. If 
\mbox{$\protect\displaystyle{\protect\mbox{\protect\boldmath{$v$%
}}}_{2\!}$} is small (but \mbox{$\protect\displaystyle{\protect
\mbox{\protect\boldmath{$v$}}}_{1\!}$} arbitrarily large), then 
the expression ${\protect\mbox{\protect\boldmath{$Q$}}}$ of \mbox
{Eq.~\mbox{$\protect\displaystyle\!$}}(\protect\ref{eq:Solution-%
 FinalRemainder}) is of order ${\protect\mbox{\protect\boldmath{%
$v$}}}_2^2$, and hence is negligible in the context of \mbox{Eq.%
 ~\mbox{$\protect\displaystyle\!$}}(\protect\ref{eq:Solution-Fin%
alMain}). If we are considering the continuous Thomas \mbox{}%
\protect\/{\protect\em precession\protect\/}, then we can set 
\mbox{$\protect\displaystyle{\protect\mbox{\protect\boldmath{$v$%
}}}_{1\!}={\protect\mbox{\protect\boldmath{$v$}}}$} and \mbox{$%
\protect\displaystyle{\protect\mbox{\protect\boldmath{$v$}}}_{2%
\!}=\delta{\protect\mbox{\protect\boldmath{$v$}}}_2$}. Then, to 
quantities of first order, \mbox{Eq.~\mbox{$\protect\displaystyle
\!$}}(\protect\ref{eq:Solution-SumV}) yields \setcounter{Ceqinde%
nt}{0}\protect\begin{eqnarray}\hspace{-1.3ex}&\displaystyle\delta
{\protect\mbox{\protect\boldmath{$v$}}}\equiv{\protect\mbox{%
\protect\boldmath{$v$}}}_{12\!}-{\protect\mbox{\protect\boldmath
{$v$}}}_1=\delta{\protect\mbox{\protect\boldmath{$v$}}}_2-({%
\protect\mbox{\protect\boldmath{$v$}}}\mbox{$\hspace{0.2ex}\cdot
\hspace{0.2ex}$}\delta{\protect\mbox{\protect\boldmath{$v$}}}_2)%
{\protect\mbox{\protect\boldmath{$v$}}}.\protect\nonumber
\setlength{\Cscr}{\value{CEht}\Ctenthex}\addtolength{\Cscr}{-1.0%
ex}\protect\raisebox{0ex}[\value{CEht}\Ctenthex][\Cscr]{}\protect
\end{eqnarray}\setcounter{CEht}{10}Thus, if $\delta{\protect\mbox
{\protect\boldmath{$v$}}}_2$ is perpendicular to the velocity ${%
\protect\mbox{\protect\boldmath{$v$}}}$, then \mbox{$\protect
\displaystyle\delta{\protect\mbox{\protect\boldmath{$v$}}}=\delta
{\protect\mbox{\protect\boldmath{$v$}}}_2$}; but if $\delta{%
\protect\mbox{\protect\boldmath{$v$}}}_2$ is parallel to ${%
\protect\mbox{\protect\boldmath{$v$}}}$, then one must take into 
account the fact that the velocity must remain smaller than that 
of light. On the other hand, in all cases we have \mbox{$\protect
\displaystyle{\protect\mbox{\protect\boldmath{$v$}}}\mbox{$\times
$}\delta{\protect\mbox{\protect\boldmath{$v$}}}={\protect\mbox{%
\protect\boldmath{$v$}}}\mbox{$\times$}\delta{\protect\mbox{%
\protect\boldmath{$v$}}}_2$}, so that \mbox{Eq.~\mbox{$\protect
\displaystyle\!$}}(\protect\ref{eq:Solution-FinalMain}) yields, 
to first order, \setcounter{Ceqindent}{0}\protect\begin{eqnarray%
}\protect\left.\protect\begin{array}{rcl}\protect\displaystyle
\hspace{-1.3ex}&\protect\displaystyle\delta{\protect\mbox{%
\protect\boldmath{$r$}}}\equiv{\protect\mbox{\protect\boldmath{$%
r$}}}^{\prime\!}-{\protect\mbox{\protect\boldmath{$r$}}}=\mbox{$%
\protect\displaystyle\protect\frac{\gamma}{\gamma+1}$}({\protect
\mbox{\protect\boldmath{$v$}}}\mbox{$\times$}\delta{\protect\mbox
{\protect\boldmath{$v$}}})\mbox{$\times$}{\protect\mbox{\protect
\boldmath{$r$}}},\setlength{\Cscr}{\value{CEht}\Ctenthex}%
\addtolength{\Cscr}{-1.0ex}\protect\raisebox{0ex}[\value{CEht}%
\Ctenthex][\Cscr]{}\protect\end{array}\protect\right.\protect
\label{eq:Solution-Precession}\protect\end{eqnarray}\setcounter{%
CEht}{10}which is the standard expression for the Thomas precess%
ion.$^{\ref{cit:Jackson1999}}$ (To compare with Jackson's result 
following his (11.117), note that our $\delta{\protect\mbox{%
\protect\boldmath{$v$}}}$ is his ${\protect\it\Delta\!\:}{%
\protect\mbox{\protect\boldmath{$\beta$}}}$, and that our \mbox{%
$\protect\displaystyle{\protect\mbox{\protect\boldmath{$v$}}}%
\mbox{$\times$}\delta{\protect\mbox{\protect\boldmath{$v$}}}$} i%
s, in his notation, \mbox{$\protect\displaystyle{\protect\mbox{%
\protect\boldmath{$\beta$}}}\mbox{$\times$}{\protect\it\Delta\!%
\:}{\protect\mbox{\protect\boldmath{$\beta$}}}=\gamma^{\,}{%
\protect\mbox{\protect\boldmath{$\beta$}}}\mbox{$\times$}\delta{%
\protect\mbox{\protect\boldmath{$\beta$}}}$}.)\par If we now con%
sider the ultrarelativistic limit of \mbox{Eq.~\mbox{$\protect
\displaystyle\!$}}(\protect\ref{eq:Solution-Precession}), then w%
e find something remarkable. This limit may be taken to be defin%
ed by the relations \setcounter{Ceqindent}{0}\protect\begin{eqna%
rray}\protect\left.\protect\begin{array}{rcl}\protect
\displaystyle\hspace{-1.3ex}&\protect\displaystyle\gamma
\rightarrow\infty,\hspace{8ex}{\protect\mbox{\protect\boldmath{$%
v$}}}^{2\!}\rightarrow1,\hspace{8ex}{\protect\mbox{\protect
\boldmath{$v$}}}\mbox{$\hspace{0.2ex}\cdot\hspace{0.2ex}$}\delta
{\protect\mbox{\protect\boldmath{$v$}}}\rightarrow0,\setlength{%
\Cscr}{\value{CEht}\Ctenthex}\addtolength{\Cscr}{-1.0ex}\protect
\raisebox{0ex}[\value{CEht}\Ctenthex][\Cscr]{}\protect\end{array%
}\protect\right.\protect\label{eq:Solution-UltraRel}\protect\end
{eqnarray}\setcounter{CEht}{10}the latter two of which simply re%
flect the fact that the velocity is at all times almost the spee%
d of light, and (hence) that any changes $\delta{\protect\mbox{%
\protect\boldmath{$v$}}}$ to the velocity ${\protect\mbox{%
\protect\boldmath{$v$}}}$ must be perpendicular to~${\protect
\mbox{\protect\boldmath{$v$}}}$. In this limit, \mbox{Eq.~\mbox{%
$\protect\displaystyle\!$}}(\protect\ref{eq:Solution-Precession}%
) becomes \setcounter{Ceqindent}{0}\protect\begin{eqnarray}%
\protect\left.\protect\begin{array}{rcl}\protect\displaystyle
\hspace{-1.3ex}&\protect\displaystyle\delta{\protect\mbox{%
\protect\boldmath{$r$}}}\rightarrow({\protect\mbox{\protect
\boldmath{$v$}}}\mbox{$\times$}\delta{\protect\mbox{\protect
\boldmath{$v$}}})\mbox{$\times$}{\protect\mbox{\protect\boldmath
{$r$}}}.\setlength{\Cscr}{\value{CEht}\Ctenthex}\addtolength{%
\Cscr}{-1.0ex}\protect\raisebox{0ex}[\value{CEht}\Ctenthex][\Cscr
]{}\protect\end{array}\protect\right.\protect\label{eq:Solution-%
Rur}\protect\end{eqnarray}\setcounter{CEht}{10}Consider, now, th%
e expression \mbox{$\protect\displaystyle({\protect\mbox{\protect
\boldmath{$v$}}}\mbox{$\times$}\delta{\protect\mbox{\protect
\boldmath{$v$}}})\mbox{$\times$}{\protect\mbox{\protect\boldmath
{$v$}}}$}. By a standard three-vector identity, we have 
\setcounter{Ceqindent}{0}\protect\begin{eqnarray}\hspace{-1.3ex}%
&\displaystyle({\protect\mbox{\protect\boldmath{$v$}}}\mbox{$%
\times$}\delta{\protect\mbox{\protect\boldmath{$v$}}})\mbox{$%
\times$}{\protect\mbox{\protect\boldmath{$v$}}}\equiv{\protect
\mbox{\protect\boldmath{$v$}}}^2\delta{\protect\mbox{\protect
\boldmath{$v$}}}-({\protect\mbox{\protect\boldmath{$v$}}}\mbox{$%
\hspace{0.2ex}\cdot\hspace{0.2ex}$}\delta{\protect\mbox{\protect
\boldmath{$v$}}}){\protect\mbox{\protect\boldmath{$v$}}},\protect
\nonumber\setlength{\Cscr}{\value{CEht}\Ctenthex}\addtolength{%
\Cscr}{-1.0ex}\protect\raisebox{0ex}[\value{CEht}\Ctenthex][\Cscr
]{}\protect\end{eqnarray}\setcounter{CEht}{10}which, on account 
of the relations (\protect\ref{eq:Solution-UltraRel}), tells us 
that, in the ultrarelativistic limit, \setcounter{Ceqindent}{0}%
\protect\begin{eqnarray}\protect\left.\protect\begin{array}{rcl}%
\protect\displaystyle\hspace{-1.3ex}&\protect\displaystyle\delta
{\protect\mbox{\protect\boldmath{$v$}}}\rightarrow({\protect\mbox
{\protect\boldmath{$v$}}}\mbox{$\times$}\delta{\protect\mbox{%
\protect\boldmath{$v$}}})\mbox{$\times$}{\protect\mbox{\protect
\boldmath{$v$}}}.\setlength{\Cscr}{\value{CEht}\Ctenthex}%
\addtolength{\Cscr}{-1.0ex}\protect\raisebox{0ex}[\value{CEht}%
\Ctenthex][\Cscr]{}\protect\end{array}\protect\right.\protect
\label{eq:Solution-Vur}\protect\end{eqnarray}\setcounter{CEht}{1%
0}Comparing (\protect\ref{eq:Solution-Rur}) and (\protect\ref{eq%
:Solution-Vur}), we thus find that ${\protect\mbox{\protect
\boldmath{$r$}}}$ and ${\protect\mbox{\protect\boldmath{$v$}}}$ 
are being rotated by the same amount about the same axis. Recall%
ing our discussion above that the direction of the Thomas rotati%
on of the \mbox{}\protect\/{\protect\em axes\protect\/} is oppos%
ite to the rotation of ${\protect\mbox{\protect\boldmath{$r$}}}$ 
\mbox{}\protect\/{\protect\em relative\protect\/} to these axes, 
we therefore find that we have proved the following remarkable t%
heorem: \mbox{}\protect\/{\protect\em For any ultrarelativistic 
object, the Thomas rotation is equal and opposite to the orbital 
rotation.\protect\/}\par This theorem explains why we obtained a 
rotation angle of \mbox{$\protect\displaystyle90\mbox{$^\circ$}{%
}^{\!}$} for our sequence of four boosts in the ultrarelativisti%
c limit. For we can think of any \mbox{}\protect\/{\protect\em f%
inite\protect\/} boost as simply a sequence of infinitesimal boo%
sts in the same direction. For our first ($+x$) boost, we simply 
boosted the \mbox{}\protect\/{\protect\em Enterprise\protect\/}'%
s velocity to ultrarelativistic speeds. The second ($+y$) boost 
was designed to bring the \mbox{}\protect\/{\protect\em Enterpri%
se\protect\/}'s velocity around to a \mbox{$\protect\displaystyle
45\mbox{$^\circ$}{}^{\!}$} angle between the $+x$ and $+y$ direc%
tions; and the third ($-x$) boost to bring it around another 
\mbox{$\protect\displaystyle45\mbox{$^\circ$}{}^{\!}$} to the $+%
y$ direction. The final ($-y$) boost was antiparallel to this ve%
locity, and simply brought the \mbox{}\protect\/{\protect\em Ent%
erprise\protect\/}\ back to rest. Thus, the velocity of the \mbox
{}\protect\/{\protect\em Enterprise\protect\/}, relative to a fi%
xed observer, was rotated by \mbox{$\protect\displaystyle90\mbox
{$^\circ$}{}^{\!}$} at ultrarelativistic speeds; and hence, by t%
he above theorem, the Thomas rotation is just $90\mbox{$^\circ$}%
$\mbox{$\!$}, which is what we found by elementary means above.%
\par\refstepcounter{section}\vspace{1.5\baselineskip}\par{%
\centering\bf\thesection. An advanced ``paradox'': polarization 
properties of scattering events\\*[0.5\baselineskip]}\protect
\indent\label{sect:Scatter}Let us now consider a more advanced s%
ituation: the calculation of a polarized cross-section in quantu%
m field theory. For simplicity, let us consider the scattering o%
f a Dirac electron by the (idealized) fixed Coulomb field of an 
infinitely heavy, pointlike nucleus. For definiteness, we shall 
follow the notation and conventions employed in the introductory 
textbook by Mandl and Shaw.$^{\ref{cit:Mandl1984}}$ In any frame 
for which the scattered electron momentum \mbox{$\protect
\displaystyle{\protect\mbox{\protect\boldmath{$p$}}}{^\prime\!}$%
} has the same magnitude as the incident momentum ${\protect\mbox
{\protect\boldmath{$p$}}}$ (i.e., for which the electron's energ%
y is unchanged by the scattering), the fully polarized cross-sec%
tion is given by$^{\ref{cit:Mandl1984}}$ \setcounter{Ceqindent}{%
0}\protect\begin{eqnarray}\protect\left.\protect\begin{array}{rc%
l}\protect\displaystyle\hspace{-1.3ex}&\protect\displaystyle
\protect\left(\mbox{$\protect\displaystyle\protect\frac{d\sigma}%
{d{\protect\it\Omega\!\:}^{\prime\!}}$}\protect\right)_{\!\!rs}%
\!\!=\protect\left(\mbox{$\protect\displaystyle\protect\frac{me}%
{2\pi}$}\protect\right)^{\!\!2}\!|{\cal M}_{rs}|^2=\protect\left
(\mbox{$\protect\displaystyle\protect\frac{me}{2\pi}$}\protect
\right)^{\!\!2}\!|\bar u_s({\protect\mbox{\protect\boldmath{$p$}%
}}')\mbox{$A\hspace{-1.15ex}\hspace{-0.1ex}\raisebox{0.2ex}\slash
\hspace{--0.1ex}\hspace{0.03ex}$}_e({\protect\mbox{\protect
\boldmath{$q$}}})u_r({\protect\mbox{\protect\boldmath{$p$}}})|^2%
\mbox{$\!$},\setlength{\Cscr}{\value{CEht}\Ctenthex}\addtolength
{\Cscr}{-1.0ex}\protect\raisebox{0ex}[\value{CEht}\Ctenthex][%
\Cscr]{}\protect\end{array}\protect\right.\protect\label{eq:Scat%
ter-XsectGen}\protect\end{eqnarray}\setcounter{CEht}{10}where $m%
$ is the mass and $-e$ the charge of the electron, ${\cal M}_{rs%
}$ is the Feynman amplitude for the process, \mbox{$\protect
\displaystyle{\protect\mbox{\protect\boldmath{$q$}}}\equiv{%
\protect\mbox{\protect\boldmath{$p$}}}^{\prime\!}-{\protect\mbox
{\protect\boldmath{$p$}}}$} is the momentum transfer, $A_e({%
\protect\mbox{\protect\boldmath{$q$}}})$ is the ``external'' ele%
ctromagnetic field (i.e., the Coulomb field of the nucleus) in m%
omentum space, and \mbox{$\protect\displaystyle\mbox{$A\hspace{-%
1.15ex}\hspace{-0.1ex}\raisebox{0.2ex}\slash\hspace{--0.1ex}%
\hspace{0.03ex}$}_e\equiv A_e^\mu\gamma_\mu$} where $\gamma_\mu$ 
are the Dirac gamma matrices (not to be confused with the factor 
$\gamma$ defined in \mbox{Eq.~\mbox{$\protect\displaystyle\!$}}(%
\protect\ref{eq:Prelim-Gamma})). The indices $r$ and $s$ (\mbox{%
$\protect\displaystyle=1,2$}) label the two possible spin states 
of the incident and scattered electron respectively.\par Let us 
first calculate all of the polarized cross-sections for the foll%
owing scenario. We choose an inertial frame in which the nucleus 
is at rest, so that$^{\ref{cit:Mandl1984}}$ \setcounter{Ceqinden%
t}{0}\protect\begin{eqnarray}\hspace{-1.3ex}&\displaystyle A_e^%
\mu({\protect\mbox{\protect\boldmath{$x$}}})=\protect\left(\mbox
{$\protect\displaystyle\protect\frac{Ze}{4\pi|{\protect\mbox{%
\protect\boldmath{$x$}}}|}$},0,0,0\protect\right)\mbox{$\!$},%
\protect\nonumber\setlength{\Cscr}{\value{CEht}\Ctenthex}%
\addtolength{\Cscr}{-1.0ex}\protect\raisebox{0ex}[\value{CEht}%
\Ctenthex][\Cscr]{}\protect\end{eqnarray}\setcounter{CEht}{10}wh%
ich under a Fourier transform yields \setcounter{Ceqindent}{0}%
\protect\begin{eqnarray}\hspace{-1.3ex}&\displaystyle A_e^\mu({%
\protect\mbox{\protect\boldmath{$q$}}})=\protect\left(\mbox{$%
\protect\displaystyle\protect\frac{Ze}{|{\protect\mbox{\protect
\boldmath{$q$}}}|^{2\!}}$},0,0,0\protect\right)\mbox{$\!$},%
\protect\nonumber\setlength{\Cscr}{\value{CEht}\Ctenthex}%
\addtolength{\Cscr}{-1.0ex}\protect\raisebox{0ex}[\value{CEht}%
\Ctenthex][\Cscr]{}\protect\end{eqnarray}\setcounter{CEht}{10}wh%
ere $Ze$ is the charge of the nucleus. We then have$^{\ref{cit:M%
andl1984}}$ \setcounter{Ceqindent}{0}\protect\begin{eqnarray}%
\protect\left.\protect\begin{array}{rcl}\protect\displaystyle
\hspace{-1.3ex}&\protect\displaystyle\protect\left(\mbox{$%
\protect\displaystyle\protect\frac{d\sigma}{d{\protect\it\Omega
\!\:}^{\prime\!}}$}\protect\right)_{\!\!rs}\!\!=\mbox{$\protect
\displaystyle\protect\frac{(2m\alpha Z)^{2\!}}{|{\protect\mbox{%
\protect\boldmath{$q$}}}|^4}$}^{\,}|\bar u_s({\protect\mbox{%
\protect\boldmath{$p$}}}'){\gamma}{}^{\raisebox{-0.25ex}{$%
\scriptstyle0$}}u_r({\protect\mbox{\protect\boldmath{$p$}}})|^2%
\mbox{$\!$},\setlength{\Cscr}{\value{CEht}\Ctenthex}\addtolength
{\Cscr}{-1.0ex}\protect\raisebox{0ex}[\value{CEht}\Ctenthex][%
\Cscr]{}\protect\end{array}\protect\right.\protect\label{eq:Scat%
ter-Xsect1}\protect\end{eqnarray}\setcounter{CEht}{10}where \mbox
{$\protect\displaystyle\alpha\equiv e^2\!/4\pi$} is the fine-str%
ucture constant. Let us consider the case when the incident elec%
tron has velocity components \setcounter{Ceqindent}{0}\protect
\begin{eqnarray}\hspace{-1.3ex}&\displaystyle v_x=-\mbox{$%
\protect\displaystyle\protect\frac{2}{3}$},\hspace{10ex}v_y=0,%
\hspace{10ex}v_z=+\mbox{$\protect\displaystyle\protect\frac{2}{3%
}$},\protect\nonumber\setlength{\Cscr}{\value{CEht}\Ctenthex}%
\addtolength{\Cscr}{-1.0ex}\protect\raisebox{0ex}[\value{CEht}%
\Ctenthex][\Cscr]{}\protect\end{eqnarray}\setcounter{CEht}{10}an%
d the scattered electron has velocity components \setcounter{Ceq%
indent}{0}\protect\begin{eqnarray}\hspace{-1.3ex}&\displaystyle
v_x=+\mbox{$\protect\displaystyle\protect\frac{2}{3}$},\hspace{1%
0ex}v_y=0,\hspace{10ex}v_z=+\mbox{$\protect\displaystyle\protect
\frac{2}{3}$},\protect\nonumber\setlength{\Cscr}{\value{CEht}%
\Ctenthex}\addtolength{\Cscr}{-1.0ex}\protect\raisebox{0ex}[%
\value{CEht}\Ctenthex][\Cscr]{}\protect\end{eqnarray}\setcounter
{CEht}{10}so that the electron is being scattered by $90\mbox{$^%
\circ$}$ in the $z$--$x$ plane. {}From\ \mbox{Eq.~\mbox{$\protect
\displaystyle\!$}}(\protect\ref{eq:Prelim-Gamma}) we find that 
\mbox{$\protect\displaystyle\gamma=3$}, so that the incident and 
scattered four-momenta have the components \setcounter{Ceqindent%
}{0}\protect\begin{eqnarray}\hspace{-1.3ex}&\displaystyle p^{\:%
\!\mu\!}=\left(\begin{array}{c}3m\\\!\!-2m\\0\\2m\end{array}%
\right)\mbox{$\!$},\hspace{12ex}p^{\prime\mu\!}=\left(\begin{arr%
ay}{c}3m\\2m\\0\\2m\end{array}\right)\mbox{$\!$},\protect
\nonumber\setlength{\Cscr}{\value{CEht}\Ctenthex}\addtolength{%
\Cscr}{-1.0ex}\protect\raisebox{0ex}[\value{CEht}\Ctenthex][\Cscr
]{}\protect\end{eqnarray}\setcounter{CEht}{10}and \mbox{$\protect
\displaystyle|{\protect\mbox{\protect\boldmath{$q$}}}|=4m$}. Now%
, the positive-energy spin--momentum eigenstates are, in the Dir%
ac--Pauli representation of the Dirac matrices, given by$^{\ref{%
cit:Mandl1984}}$ \setcounter{Ceqindent}{0}\protect\begin{eqnarra%
y}\protect\left.\protect\begin{array}{rcl}\protect\displaystyle
\hspace{-1.3ex}&\protect\displaystyle u_1({\protect\mbox{\protect
\boldmath{$p$}}})=c_{1\!}\!\left(\begin{array}{c}1\\0\\c_2p^{\:%
\!z}\\c_2(p^{\:\!x\!}+ip^{\:\!y})\end{array}\right)\mbox{$\!$},%
\hspace{10ex}u_2({\protect\mbox{\protect\boldmath{$p$}}})=c_{1\!%
}\!\left(\begin{array}{c}0\\1\\c_2(p^{\:\!x\!}-ip^{\:\!y})\\\!\!%
-c_2p^{\:\!z}\end{array}\right)\mbox{$\!$},\setlength{\Cscr}{%
\value{CEht}\Ctenthex}\addtolength{\Cscr}{-1.0ex}\protect
\raisebox{0ex}[\value{CEht}\Ctenthex][\Cscr]{}\protect\end{array%
}\protect\right.\protect\label{eq:Scatter-Bispinors}\protect\end
{eqnarray}\setcounter{CEht}{10}where \setcounter{Ceqindent}{0}%
\protect\begin{eqnarray}\protect\left.\protect\begin{array}{rcl}%
\protect\displaystyle\hspace{-1.3ex}&\protect\displaystyle c_1%
\equiv\sqrt{\mbox{$\protect\displaystyle\protect\frac{E+m}{2m}$}%
},\hspace{12ex}c_2\equiv\mbox{$\protect\displaystyle\protect\frac
{1}{E+m}$},\setlength{\Cscr}{\value{CEht}\Ctenthex}\addtolength{%
\Cscr}{-1.0ex}\protect\raisebox{0ex}[\value{CEht}\Ctenthex][\Cscr
]{}\protect\end{array}\protect\right.\protect\label{eq:Scatter-C%
12}\protect\end{eqnarray}\setcounter{CEht}{10}where \mbox{$%
\protect\displaystyle u_{1\!}$} ($u_2$) is the spin-up (spin-dow%
n) eigenstate relative to the $z$-direction. The conjugate bispi%
nor eigenstates, in this representation, are consequently 
\setcounter{Ceqindent}{0}\protect\begin{eqnarray}\protect\left.%
\protect\begin{array}{rcl}\protect\displaystyle\hspace{-1.3ex}&%
\protect\displaystyle\bar u_1({\protect\mbox{\protect\boldmath{$%
p$}}})=c_{1\!}\!\left(\begin{array}{c}1\\0\\\!\!-c_2p^{\:\!z}\\c%
_2(-p^{\:\!x\!}+ip^{\:\!y})\end{array}\right)^{\mbox{\scriptsize
$\!\!$T}}\!\!,\hspace{10ex}\bar u_2({\protect\mbox{\protect
\boldmath{$p$}}})=c_{1\!}\!\left(\begin{array}{c}0\\1\\\!\!-c_2(%
p^{\:\!x\!}+ip^{\:\!y})\\c_2p^{\:\!z}\end{array}\right)^{\mbox{%
\scriptsize$\!\!$T}}\!\!.\setlength{\Cscr}{\value{CEht}\Ctenthex
}\addtolength{\Cscr}{-1.0ex}\protect\raisebox{0ex}[\value{CEht}%
\Ctenthex][\Cscr]{}\protect\end{array}\protect\right.\protect
\label{eq:Scatter-ConjBispinors}\protect\end{eqnarray}\setcounter
{CEht}{10}For our particular case, \mbox{$\protect\displaystyle
E=3m$}, so \mbox{$\protect\displaystyle c_{1\!}=\sqrt2$} and 
\mbox{$\protect\displaystyle c_{2\!}=1/4m$}. For the incident el%
ectron we therefore have \setcounter{Ceqindent}{0}\protect\begin
{eqnarray}\hspace{-1.3ex}&\displaystyle u_1({\protect\mbox{%
\protect\boldmath{$p$}}})=\mbox{$\protect\displaystyle\protect
\frac{1}{\sqrt2}$}\!\left(\begin{array}{c}2\\0\\1\\\!\!-1\!\end{%
array}\right)\mbox{$\!$},\hspace{10ex}u_2({\protect\mbox{\protect
\boldmath{$p$}}})=\mbox{$\protect\displaystyle\protect\frac{1}{%
\sqrt2}$}\!\left(\begin{array}{c}0\\2\\\!\!-1\!\\\!\!-1\!\end{ar%
ray}\right)\mbox{$\!$},\protect\nonumber\setlength{\Cscr}{\value
{CEht}\Ctenthex}\addtolength{\Cscr}{-1.0ex}\protect\raisebox{0ex%
}[\value{CEht}\Ctenthex][\Cscr]{}\protect\end{eqnarray}%
\setcounter{CEht}{10}and for the scattered electron we have 
\setcounter{Ceqindent}{0}\protect\begin{eqnarray}\hspace{-1.3ex}%
&\displaystyle\bar u_1({\protect\mbox{\protect\boldmath{$p$}}}')%
=\mbox{$\protect\displaystyle\protect\frac{1}{\sqrt2}$}\!\left(%
\begin{array}{c}2\\0\\\!\!-1\!\\\!\!-1\!\end{array}\right)^{\mbox
{\scriptsize$\!\!$T}}\!\!,\hspace{10ex}\bar u_2({\protect\mbox{%
\protect\boldmath{$p$}}}')=\mbox{$\protect\displaystyle\protect
\frac{1}{\sqrt2}$}\!\left(\begin{array}{c}0\\2\\\!\!-1\!\\1\end{%
array}\right)^{\mbox{\scriptsize$\!\!$T}}\!\!.\protect\nonumber
\setlength{\Cscr}{\value{CEht}\Ctenthex}\addtolength{\Cscr}{-1.0%
ex}\protect\raisebox{0ex}[\value{CEht}\Ctenthex][\Cscr]{}\protect
\end{eqnarray}\setcounter{CEht}{10}Let us now compute the cross-%
section (\protect\ref{eq:Scatter-Xsect1}). The quantity \mbox{$%
\protect\displaystyle\bar u_s({\protect\mbox{\protect\boldmath{$%
p$}}}'){\gamma}{}^{\raisebox{-0.25ex}{$\scriptstyle0$}}u_r({%
\protect\mbox{\protect\boldmath{$p$}}})$} is equal to 2 for no s%
pin flip in the $z$-direction (i.e., for \mbox{$\protect
\displaystyle\bar u_{1\!}$} with $u_1$, or for \mbox{$\protect
\displaystyle\bar u_{2\!}$} with $u_2$), and is equal to $\pm1$ 
for spin flip in the $z$-direction (i.e., for \mbox{$\protect
\displaystyle\bar u_{1\!}$} with $u_2$, or for \mbox{$\protect
\displaystyle\bar u_{2\!}$} with $u_1$). We thus find that 
\setcounter{Ceqindent}{0}\protect\begin{eqnarray}\protect\left.%
\protect\begin{array}{rcl}\protect\displaystyle\hspace{-1.3ex}&%
\protect\displaystyle\mbox{$\protect\displaystyle\protect\frac{d%
\sigma}{d{\protect\it\Omega\!\:}^{\prime\!}}$}^{\:\!}(\mbox{no s%
pin flip})=\mbox{$\protect\displaystyle\protect\frac{\alpha^{2\!%
}Z^2}{16m^2}$},\hspace{10ex}\mbox{$\protect\displaystyle\protect
\frac{d\sigma}{d{\protect\it\Omega\!\:}^{\prime\!}}$}^{\:\!}(%
\mbox{spin flip})=\mbox{$\protect\displaystyle\protect\frac{%
\alpha^{2\!}Z^2}{64m^2}$}.\setlength{\Cscr}{\value{CEht}\Ctenthex
}\addtolength{\Cscr}{-1.0ex}\protect\raisebox{0ex}[\value{CEht}%
\Ctenthex][\Cscr]{}\protect\end{array}\protect\right.\protect
\label{eq:Scatter-Pol1}\protect\end{eqnarray}\setcounter{CEht}{1%
0}To obtain the unpolarized cross-section, we average over the i%
nitial spin states and sum over the final spin states in the sta%
ndard way. This results in \setcounter{Ceqindent}{0}\protect
\begin{eqnarray}\hspace{-1.3ex}&\displaystyle\mbox{$\protect
\displaystyle\protect\frac{d\sigma}{d{\protect\it\Omega\!\:}^{%
\prime\!}}$}^{\:\!}(\mbox{unpolarized})\equiv\mbox{$\protect
\displaystyle\protect\frac{1}{2}$}\sum_{r=1}^2\sum_{s=1}^2\!%
\protect\left(\mbox{$\protect\displaystyle\protect\frac{d\sigma}%
{d{\protect\it\Omega\!\:}^{\prime\!}}$}\protect\right)_{\!\!rs}%
\!\!=\mbox{$\protect\displaystyle\protect\frac{d\sigma}{d{%
\protect\it\Omega\!\:}^{\prime\!}}$}^{\:\!}(\mbox{no spin flip})%
+\mbox{$\protect\displaystyle\protect\frac{d\sigma}{d{\protect\it
\Omega\!\:}^{\prime\!}}$}^{\:\!}(\mbox{spin flip})=\mbox{$%
\protect\displaystyle\protect\frac{5\alpha^{2\!}Z^2}{64m^2}$}.%
\protect\nonumber\setlength{\Cscr}{\value{CEht}\Ctenthex}%
\addtolength{\Cscr}{-1.0ex}\protect\raisebox{0ex}[\value{CEht}%
\Ctenthex][\Cscr]{}\protect\end{eqnarray}\setcounter{CEht}{10}We 
can compare this result with the standard Mott scattering formul%
a,$^{\ref{cit:Mandl1984}}$ \setcounter{Ceqindent}{0}\protect
\begin{eqnarray}\hspace{-1.3ex}&\displaystyle\mbox{$\protect
\displaystyle\protect\frac{d\sigma}{d{\protect\it\Omega\!\:}^{%
\prime\!}}$}^{\:\!}(\mbox{Mott})=\mbox{$\protect\displaystyle
\protect\frac{(\alpha Z)^2}{4E^2v^4\sin^4(\theta/2)}$}[1-v^2\sin
^2(\theta/2)],\protect\nonumber\setlength{\Cscr}{\value{CEht}%
\Ctenthex}\addtolength{\Cscr}{-1.0ex}\protect\raisebox{0ex}[%
\value{CEht}\Ctenthex][\Cscr]{}\protect\end{eqnarray}\setcounter
{CEht}{10}by noting that, for our case, \mbox{$\protect
\displaystyle\theta=90\mbox{$^\circ$}$} so \mbox{$\protect
\displaystyle\theta/2=45\mbox{$^\circ$}$} and hence \mbox{$%
\protect\displaystyle\sin^2(\theta/2)=1/2$}; \mbox{$\protect
\displaystyle v^{2\!}=8/9$}; and \mbox{$\protect\displaystyle E=%
3m$}; which yields precisely the same result: \setcounter{Ceqind%
ent}{0}\protect\begin{eqnarray}\hspace{-1.3ex}&\displaystyle\mbox
{$\protect\displaystyle\protect\frac{d\sigma}{d{\protect\it\Omega
\!\:}^{\prime\!}}$}^{\:\!}(\mbox{Mott})=\mbox{$\protect
\displaystyle\protect\frac{5\alpha^{2\!}Z^2}{64m^2}$}.\protect
\nonumber\setlength{\Cscr}{\value{CEht}\Ctenthex}\addtolength{%
\Cscr}{-1.0ex}\protect\raisebox{0ex}[\value{CEht}\Ctenthex][\Cscr
]{}\protect\end{eqnarray}\setcounter{CEht}{10}We may therefore b%
e confident that we have not made any elementary mistakes in cal%
culating the polarized cross-sections (\protect\ref{eq:Scatter-P%
ol1}).\par Let us now compute these cross-sections from the poin%
t of view of a different inertial frame. Specifically, let us vi%
ew the process from an inertial frame which moves along the posi%
tive $z$-axis with velocity $2/3$ relative to the inertial frame 
used above. Applying $B_z(2/3)$ to the components of \mbox{$%
\protect\displaystyle p^{\:\!\mu\!}$} and \mbox{$\protect
\displaystyle p^{\prime\:\!\mu}$}\mbox{$\!$}, we find \setcounter
{Ceqindent}{0}\protect\begin{eqnarray}\protect\left.\protect
\begin{array}{rcl}\protect\displaystyle\hspace{-1.3ex}&\protect
\displaystyle B_z(2/3)^{\!}\left(\begin{array}{c}3m\\\!\!-2m\\0%
\\2m\end{array}\right)=\left(\begin{array}{c}m\sqrt5\\\!\!-2m\\0%
\\0\end{array}\right)\mbox{$\!$},\hspace{8ex}B_z(2/3)^{\!}\left(%
\begin{array}{c}3m\\2m\\0\\2m\end{array}\right)=\left(\begin{arr%
ay}{c}m\sqrt5\\2m\\0\\0\end{array}\right)\mbox{$\!$},\setlength{%
\Cscr}{\value{CEht}\Ctenthex}\addtolength{\Cscr}{-1.0ex}\protect
\raisebox{0ex}[\value{CEht}\Ctenthex][\Cscr]{}\protect\end{array%
}\protect\right.\protect\label{eq:Scatter-BoostedP}\protect\end{%
eqnarray}\setcounter{CEht}{10}so that, from the point of view of 
this new frame, the electron travels in the negative-$x$ directi%
on with energy \mbox{$\protect\displaystyle E=m\sqrt5$} and spee%
d $2/\sqrt5$, and is then reflected elastically to travel in the 
positive-$x$ direction with the same energy and speed. We also n%
eed to boost the components of the four-potential $A_e^\mu$: 
\setcounter{Ceqindent}{0}\protect\begin{eqnarray}\hspace{-1.3ex}%
&\displaystyle B_z(2/3)^{\,}A_e^\mu({\protect\mbox{\protect
\boldmath{$q$}}})=\mbox{$\protect\displaystyle\protect\frac{Ze}{%
\sqrt5|{\protect\mbox{\protect\boldmath{$q$}}}|^{2\!}}$}^{\,}(3,%
0,0,-2),\protect\nonumber\setlength{\Cscr}{\value{CEht}\Ctenthex
}\addtolength{\Cscr}{-1.0ex}\protect\raisebox{0ex}[\value{CEht}%
\Ctenthex][\Cscr]{}\protect\end{eqnarray}\setcounter{CEht}{10}so 
that the equivalent expression to (\protect\ref{eq:Scatter-Xsect%
1}) for the polarized cross-section is \setcounter{Ceqindent}{0}%
\protect\begin{eqnarray}\protect\left.\protect\begin{array}{rcl}%
\protect\displaystyle\hspace{-1.3ex}&\protect\displaystyle
\protect\left(\mbox{$\protect\displaystyle\protect\frac{d\sigma}%
{d{\protect\it\Omega\!\:}^{\prime\!}}$}\protect\right)_{\!\!rs}%
\!\!=\mbox{$\protect\displaystyle\protect\frac{(2m\alpha Z)^{2\!%
}}{5|{\protect\mbox{\protect\boldmath{$q$}}}|^4}$}^{\,}|3\bar u_%
s({\protect\mbox{\protect\boldmath{$p$}}}'){\gamma}{}^{\raisebox
{-0.25ex}{$\scriptstyle0$}}u_r({\protect\mbox{\protect\boldmath{%
$p$}}})+2\bar u_s({\protect\mbox{\protect\boldmath{$p$}}}'){%
\gamma}{}^{\raisebox{-0.25ex}{$\scriptstyle z$}}u_r({\protect
\mbox{\protect\boldmath{$p$}}})|^2\mbox{$\!$}.\setlength{\Cscr}{%
\value{CEht}\Ctenthex}\addtolength{\Cscr}{-1.0ex}\protect
\raisebox{0ex}[\value{CEht}\Ctenthex][\Cscr]{}\protect\end{array%
}\protect\right.\protect\label{eq:Scatter-Xsect2}\protect\end{eq%
narray}\setcounter{CEht}{10}(We would, in general, need to trans%
form the argument ${\protect\mbox{\protect\boldmath{$q$}}}$ as w%
ell as the components $A_e^\mu$ under a Lorentz transformation. 
However, if we define \mbox{$\protect\displaystyle q^{\mu\!}%
\equiv p^{\prime\mu\!}-p^{\:\!\mu}$}\mbox{$\!$}, then in the ori%
ginal frame \mbox{$\protect\displaystyle|{\protect\mbox{\protect
\boldmath{$q$}}}|^2=-q^\mu q_\mu$} because \mbox{$\protect
\displaystyle q^{0\!}=0$}, i.e., the electron energy is conserve%
d. Since \mbox{$\protect\displaystyle q^\mu q_\mu$} is a Lorentz 
scalar, then we find that $|{\protect\mbox{\protect\boldmath{$q$%
}}}|^2$ is invariant in any frame in which the electron energy i%
s conserved---as is the case in the frame we have defined above.%
) Finally, from \mbox{Eqs.~\mbox{$\protect\displaystyle\!$}}(%
\protect\ref{eq:Scatter-C12}) we find that, using the boosted mo%
mentum values (\protect\ref{eq:Scatter-BoostedP}), the constants 
\mbox{$\protect\displaystyle c_{1\!}$} and \mbox{$\protect
\displaystyle c_{2\!}$} are given by \setcounter{Ceqindent}{0}%
\protect\begin{eqnarray}\hspace{-1.3ex}&\displaystyle c_{1\!}=%
\sqrt{\mbox{$\protect\displaystyle\protect\frac{\raisebox{0ex}[1%
.5ex][0ex]{$1+^{\!}\sqrt5$}}{2}$}},\hspace{24ex}c_{2\!}=\mbox{$%
\protect\displaystyle\protect\frac{1}{m(1+^{\!}\sqrt5)}$},%
\protect\nonumber\setlength{\Cscr}{\value{CEht}\Ctenthex}%
\addtolength{\Cscr}{-1.0ex}\protect\raisebox{0ex}[\value{CEht}%
\Ctenthex][\Cscr]{}\protect\end{eqnarray}\setcounter{CEht}{10}so 
that for the incident electron we have \setcounter{Ceqindent}{0}%
\protect\begin{eqnarray}\hspace{-1.3ex}&\displaystyle u_1({%
\protect\mbox{\protect\boldmath{$p$}}})=\mbox{$\protect
\displaystyle\protect\frac{1}{\sqrt{\raisebox{0ex}[2ex][0ex]{$2(%
1+^{\!}\sqrt5$)}}}$}\!\left(\begin{array}{c}1+^{\!}\sqrt5\\0\\0%
\\\!\!-2\!\end{array}\right)\mbox{$\!$},\hspace{6ex}u_2({\protect
\mbox{\protect\boldmath{$p$}}})=\mbox{$\protect\displaystyle
\protect\frac{1}{\sqrt{\raisebox{0ex}[2ex][0ex]{$2(1+^{\!}\sqrt5%
$)}}}$}\!\left(\begin{array}{c}0\\1+^{\!}\sqrt5\\\!\!-2\!\\0\end
{array}\right)\mbox{$\!$},\protect\nonumber\setlength{\Cscr}{%
\value{CEht}\Ctenthex}\addtolength{\Cscr}{-1.0ex}\protect
\raisebox{0ex}[\value{CEht}\Ctenthex][\Cscr]{}\protect\end{eqnar%
ray}\setcounter{CEht}{10}and for the scattered electron we have 
\setcounter{Ceqindent}{0}\protect\begin{eqnarray}\hspace{-1.3ex}%
&\displaystyle\bar u_1({\protect\mbox{\protect\boldmath{$p$}}}')%
=\mbox{$\protect\displaystyle\protect\frac{1}{\sqrt{\raisebox{0e%
x}[2ex][0ex]{$2(1+^{\!}\sqrt5$)}}}$}\!\left(\begin{array}{c}1+^{%
\!}\sqrt5\\0\\0\\\!\!-2\!\end{array}\right)^{\mbox{\scriptsize$%
\!\!$T}}\!\!,\hspace{6ex}\bar u_2({\protect\mbox{\protect
\boldmath{$p$}}}')=\mbox{$\protect\displaystyle\protect\frac{1}{%
\sqrt{\raisebox{0ex}[2ex][0ex]{$2(1+^{\!}\sqrt5$)}}}$}\!\left(%
\begin{array}{c}0\\1+^{\!}\sqrt5\\\!\!-2\!\\0\end{array}\right)^%
{\mbox{\scriptsize$\!\!$T}}\!\!.\protect\nonumber\setlength{\Cscr
}{\value{CEht}\Ctenthex}\addtolength{\Cscr}{-1.0ex}\protect
\raisebox{0ex}[\value{CEht}\Ctenthex][\Cscr]{}\protect\end{eqnar%
ray}\setcounter{CEht}{10}We now find that the quantity \mbox{$%
\protect\displaystyle\bar u_s({\protect\mbox{\protect\boldmath{$%
p$}}}'){\gamma}{}^{\raisebox{-0.25ex}{$\scriptstyle0$}}u_r({%
\protect\mbox{\protect\boldmath{$p$}}})$} is unity for no spin f%
lip in the $z$-direction, but vanishes for spin flip. The quanti%
ty \mbox{$\protect\displaystyle\bar u_s({\protect\mbox{\protect
\boldmath{$p$}}}'){\gamma}{}^{\raisebox{-0.25ex}{$\scriptstyle z%
$}}u_r({\protect\mbox{\protect\boldmath{$p$}}})$}, on the other 
hand, vanishes for no spin flip, but has the value $\pm2$ for sp%
in flip. Inserting these values into the expression (\protect\ref
{eq:Scatter-Xsect2}), we find that \setcounter{Ceqindent}{0}%
\protect\begin{eqnarray}\protect\left.\protect\begin{array}{rcl}%
\protect\displaystyle\hspace{-1.3ex}&\protect\displaystyle\mbox{%
$\protect\displaystyle\protect\frac{d\sigma}{d{\protect\it\Omega
\!\:}^{\prime\!}}$}^{\:\!}(\mbox{no spin flip})=\mbox{$\protect
\displaystyle\protect\frac{9\alpha^{2\!}Z^2}{320m^2}$},\hspace{1%
0ex}\mbox{$\protect\displaystyle\protect\frac{d\sigma}{d{\protect
\it\Omega\!\:}^{\prime\!}}$}^{\:\!}(\mbox{spin flip})=\mbox{$%
\protect\displaystyle\protect\frac{\alpha^{2\!}Z^2}{20m^2}$}.%
\setlength{\Cscr}{\value{CEht}\Ctenthex}\addtolength{\Cscr}{-1.0%
ex}\protect\raisebox{0ex}[\value{CEht}\Ctenthex][\Cscr]{}\protect
\end{array}\protect\right.\protect\label{eq:Scatter-Pol2}\protect
\end{eqnarray}\setcounter{CEht}{10}We've struck another disaster%
! The co\-efficients $9/320$ and $1/20$ in (\protect\ref{eq:Scat%
ter-Pol2}) look nothing at all like the values $1/16$ and $1/64$ 
that we found in (\protect\ref{eq:Scatter-Pol1}). But we have me%
rely performed the \mbox{}\protect\/{\protect\em same\protect\/} 
calculation in two different inertial frames! How on Earth could 
the value of the cross-section---which can be directly related t%
o the number of particles that would be expected to be measured 
in an appropriately configured experiment---depend on an arbitra%
ry choice of theoretical viewpoint? For example, if we prepare a 
beam of incident electrons so that they are completely polarized 
in the $z$-direction, and filter the scattered electrons so that 
only those polarized in the $z$-direction are detected, then wha%
t would the cross-section be: \mbox{$\protect\displaystyle\alpha
^{2\!}Z^2\!/16m^{2\!}$} or \mbox{$\protect\displaystyle9\alpha^{%
2\!}Z^2\!/320m^2$}? There cannot be two different answers!\par O%
ne might, at first glance, suspect that some trivial mistake or 
oversight has been made. However, the calculations above can be 
checked; they do not contain any arithmetical errors. Failing th%
is, one might then suspect that we have not taken into account t%
he transformation of the solid angle differential \mbox{$\protect
\displaystyle d{\protect\it\Omega\!\:}^{\prime\!}$} under a Lore%
ntz boost. However, if one checks the derivation$^{\ref{cit:Mand%
l1984}}$ of the first of the relations (\protect\ref{eq:Scatter-%
XsectGen}), then one finds that it holds true in \mbox{}\protect
\/{\protect\em any\protect\/} elastic scattering of a single par%
ticle from an ``external'' field---essentially, the other kinema%
tical factors happen to ``cancel out'' in this special class of 
scattering events.\par There is, of course, a simple way to conf%
irm or refute any suspicion one might have about the veracity of 
the results (\protect\ref{eq:Scatter-Pol2}): one need simply com%
bine them to find the \mbox{}\protect\/{\protect\em unpolarized%
\protect\/} cross-section. Surely, any trivial errors made in ob%
taining the results (\protect\ref{eq:Scatter-Pol2}) would (in al%
l but the most contrived of situations) render the unpolarized c%
ombination similarly erroneous. But we are now flabbergasted to 
find that \setcounter{Ceqindent}{0}\protect\begin{eqnarray}%
\hspace{-1.3ex}&\displaystyle\mbox{$\protect\displaystyle\protect
\frac{9}{320}$}+\mbox{$\protect\displaystyle\protect\frac{1}{20}%
$}=\mbox{$\protect\displaystyle\protect\frac{1}{16}$}+\mbox{$%
\protect\displaystyle\protect\frac{1}{64}$}=\mbox{$\protect
\displaystyle\protect\frac{5}{64}$}.\protect\nonumber\setlength{%
\Cscr}{\value{CEht}\Ctenthex}\addtolength{\Cscr}{-1.0ex}\protect
\raisebox{0ex}[\value{CEht}\Ctenthex][\Cscr]{}\protect\end{eqnar%
ray}\setcounter{CEht}{10}Thus, even though we have obtained two 
sets of irreconcilably contradictory polarized cross-sections, w%
e find that their unpolarized combinations agree completely (and 
agree with the standard Mott formula)!\par What is going on?\par
\refstepcounter{section}\vspace{1.5\baselineskip}\par{\centering
\bf\thesection. Solution to the polarization ``paradox''\\*[0.5%
\baselineskip]}\protect\indent\label{sect:ScattSol}Let us now us%
e the general discussion of \mbox{Sec.~$\:\!\!$}\protect\ref{sec%
t:Solution} to understand the polarization ``paradox'' of the pr%
evious section. The key flaw in the arguments presented above is 
the description ``polarized in the $z$-direction''\mbox{$\!$}. 
\mbox{}\protect\/{\protect\em We have not specified whose $z$-di%
rection is being used!\protect\/} The second calculation is simp%
ler, in this regard, because the electron's final velocity is co%
llinear with its initial velocity (i.e., it is in the same direc%
tion, but has the opposite sense). Thus, it is consistent for us 
to define ``the $z$-direction'' to be \mbox{}\protect\/{\protect
\em our\protect\/} $z$-axis, since all boosts to the electron's 
frames of reference are collinear. We can, say, prepare an elect%
ron polarized in the $z$-direction, and measure only those scatt%
ered electrons polarized in the $z$-direction, without ambiguity%
.\par The first calculation, on the other hand, is more subtle. 
In using the standard expressions (\protect\ref{eq:Scatter-Bispi%
nors}) and (\protect\ref{eq:Scatter-ConjBispinors}), we are (imp%
licitly) applying one single Lorentz boost from our frame of ref%
erence to the initial electron's frame, and another single Loren%
tz boost from our frame to the scattered electron's frame. These 
two boosts, however, are not collinear; and so \mbox{}\protect\/%
{\protect\em our\protect\/} description of events is different f%
rom how the \mbox{}\protect\/{\protect\em electron\protect\/} wo%
uld describe matters. (By giving the electron apparently human p%
owers, we are of course imagining an observer traveling along wi%
th the electron.) In effect, the electron's very rest frame is 
\mbox{}\protect\/{\protect\em Thomas-rotated\protect\/} by the s%
cattering event, relative to us. For example, imagine that the e%
lectron state does not get spin-flipped, as determined by the el%
ectron itself. {}From\ \mbox{}\protect\/{\protect\em our\protect
\/} point of view, however, the direction of polarization of the 
electron has changed!\par The lesson of this example is clear. I%
f one has need to calculate relativistic polarized cross-section%
s explicitly, and if the incident and scattered momenta of the p%
articles involved are not absolutely collinear (and in most prac%
tical experiments they are not), then one must be extremely caut%
ious about how one defines the spins or polarizations of the par%
ticles involved. In particular, kinematical and semi-classical a%
rguments must be examined in fine detail, to ensure that the non%
relativistic concept of universality of orientation has not been 
inappropriately applied.\par Finally, we may use the expressions 
(\protect\ref{eq:Solution-SumV}) and (\protect\ref{eq:Solution-F%
inal}) to re-analyze these polarized cross-section calculations 
\mbox{}\protect\/{\protect\em quantitatively\protect\/}. If we s%
et \mbox{$\protect\displaystyle{\protect\mbox{\protect\boldmath{%
$v$}}}_{1\!}$} to be the initial electron velocity, namely, \mbox
{$\protect\displaystyle(-2/3,0,2/3)$}, then it is straightforwar%
d to verify that a boost by \mbox{$\protect\displaystyle{\protect
\mbox{\protect\boldmath{$v$}}}_{2\!}=(12/13,0,0)$} results in th%
e correct final electron velocity of \mbox{$\protect\displaystyle
{\protect\mbox{\protect\boldmath{$v$}}}_{12\!}=(2/3,0,2/3)$}. If 
one sets ${\protect\mbox{\protect\boldmath{$r$}}}$ to be, say, 
\mbox{$\protect\displaystyle(0,0,1)$}, then, after some calculat%
ion, one finds that \mbox{$\protect\displaystyle{\protect\mbox{%
\protect\boldmath{$r$}}}^{\prime\!}=(4/5,0,3/5)$}. Thus, the ele%
ctron's rest frame has been Thomas-rotated by an angle \mbox{$%
\protect\displaystyle\theta_T=\mbox{arctan}(4/3)\approx53\mbox{$%
^\circ$}$} in the $z$--$x$ plane. If we now list the matrix elem%
ents corresponding to the polarized Feynman \mbox{}\protect\/{%
\protect\em amplitudes\protect\/} found in \mbox{Sec.~$\:\!\!$}%
\protect\ref{sect:Scatter} (rather than the cross-sections), the%
n for the first and second frames of reference we found, respect%
ively, \setcounter{Ceqindent}{0}\protect\begin{eqnarray}\hspace{%
-1.3ex}&\displaystyle{\cal M}_{rs}^{(1)}=\mbox{$\protect
\displaystyle\protect\frac{2m\alpha Z}{|{\protect\mbox{\protect
\boldmath{$q$}}}|^{2\!}}$}\!\protect\left(\begin{array}{cc}2&\!%
\!\!-1\!\\1&2\end{array}\protect\right)\!,\hspace{8ex}{\cal M}_{%
rs}^{(2)}=\mbox{$\protect\displaystyle\protect\frac{2m\alpha Z}{%
\sqrt5|{\protect\mbox{\protect\boldmath{$q$}}}|^{2\!}}$}\!%
\protect\left(\begin{array}{cc}3&\!\!\!-4\!\\4&3\end{array}%
\protect\right)\!,\protect\nonumber\setlength{\Cscr}{\value{CEht%
}\Ctenthex}\addtolength{\Cscr}{-1.0ex}\protect\raisebox{0ex}[%
\value{CEht}\Ctenthex][\Cscr]{}\protect\end{eqnarray}\setcounter
{CEht}{10}where the rows in these matrices represent the $z$-com%
ponent of the initial spin, and the columns the $z$-component of 
the final spin. To reconcile these Feynman amplitudes, we need s%
imply apply the Thomas rotation or its inverse to either the ini%
tial or the final spin state in the first frame of reference. Re%
membering that spinors transform under rotations by half-angles, 
and noting that \mbox{$\protect\displaystyle\cos(\theta_{T^{\!}}%
/2)=2/\sqrt5$} and \mbox{$\protect\displaystyle\sin(\theta_{T^{%
\!}}/2)=1/\sqrt5$}, we finally obtain \setcounter{Ceqindent}{0}%
\protect\begin{eqnarray}\hspace{-1.3ex}&\displaystyle\protect
\left(\begin{array}{cc}\cos^{\:\!\!}\mbox{$\protect\displaystyle
\protect\frac{\theta_T}{2}$}\raisebox{0ex}[3ex][3ex]{}&\!\!-^{\:%
\!\!}\sin^{\:\!\!}\mbox{$\protect\displaystyle\protect\frac{%
\theta_T}{2}$}\!\\\sin^{\:\!\!}\mbox{$\protect\displaystyle
\protect\frac{\theta_T}{2}$}\raisebox{0ex}[3ex][3ex]{}&\cos^{\:%
\!\!}\mbox{$\protect\displaystyle\protect\frac{\theta_T}{2}$}\end
{array}\protect\right)\!{\cal M}_{rs}^{(1)}={\cal M}_{rs}^{(2)\!%
}.\protect\nonumber\setlength{\Cscr}{\value{CEht}\Ctenthex}%
\addtolength{\Cscr}{-1.0ex}\protect\raisebox{0ex}[\value{CEht}%
\Ctenthex][\Cscr]{}\protect\end{eqnarray}\setcounter{CEht}{10}%
\par\refstepcounter{section}\vspace{1.5\baselineskip}\par{%
\centering\bf\thesection. Conclusions\\*[0.5\baselineskip]}%
\protect\indent\label{sect:Conclude}We have shown how the Thomas 
rotation of relativistic mechanics can be introduced, and its ``%
paradoxical'' nature discussed, at quite an introductory level; 
that resolving such ``paradoxes'' is not overly difficult; and t%
hat a general expression for arbitrary Thomas rotations can be o%
btained without excessive effort. We have also shown how this ge%
neral result connects up with standard textbook accounts of the 
infinitesimal Thomas precession. We have endeavored to show that 
the ramifications of such effects are deep, and fundamental, and 
that they may also be of immediate practical importance in the a%
nalysis and interpretation of relativistic polarized scattering 
experiments.\par In the interests of keeping this discussion at 
an introductory level, we have refrained from utilizing more adv%
anced theoretical concepts to explain or analyze the Thomas rota%
tion more elegantly or concisely. For example, group-theoretical 
methods are hinted at in the above derivations, but are not made 
explicit. (See, for example, Ref.~\ref{cit:Jackson1999} for a th%
orough treatment in these terms.) Boosts can be viewed as simply 
``rotations'' between space and time; and since two rotations ab%
out different spatial axes do not, in general, commute, then one 
would (rightly) presume that two boosts in different directions 
do not commute either; this is another path to the Thomas rotati%
on. Alternatively, one may make use of the concept of parallel t%
ransport---more familiar in the general theory of relativity, bu%
t equally applicable to boosts or accelerations in flat spacetim%
e---to arrive at the Thomas rotation by yet another path.$^{\ref
{cit:Misner1970}}$ We believe that all of these more abstract vi%
ews of the Thomas rotation do, in fact, augment, rather than det%
ract from, the elementary nature and beauty of the effect as des%
cribed here.\par\vspace{1.5\baselineskip}\par{\centering\bf Ackn%
owledgments\\*[0.5\baselineskip]}\protect\indent This work was s%
upported in part by the Australian Research Council. Helpful dis%
cussions with Brian~J.~Morphett are gratefully acknowledged. Thi%
s paper is dedicated to the memory of A.~J.~Drinan.\par\vspace{1%
.5\baselineskip}\par{\centering\bf References\\*[0.5\baselineskip
]}{\protect\mbox{}}\vspace{-\baselineskip}\vspace{-2ex}%
\settowidth\CGDnum{[\ref{citlast}]}\setlength{\CGDtext}{%
\textwidth}\addtolength{\CGDtext}{-\CGDnum}\begin{list}{Error!}{%
\setlength{\labelwidth}{\CGDnum}\setlength{\labelsep}{0.75ex}%
\setlength{\leftmargin}{0ex}\setlength{\rightmargin}{0ex}%
\setlength{\itemsep}{0ex}\setlength{\parsep}{0ex}}\protect
\frenchspacing\setcounter{CBtnc}{1}\item[{\hfill\makebox[0ex][r]%
{\raisebox{0ex}[1ex][0ex]{$^{\mbox{$\fnsymbol{CBtnc}$}}$}}}]%
\addtocounter{CBtnc}{1}Current address: Mentone Grammar, 63 Veni%
ce Street, Mentone, Victoria 3194, Australia.\item[{\hfill
\makebox[0ex][r]{\raisebox{0ex}[1ex][0ex]{$^{\mbox{$\fnsymbol{CB%
tnc}$}}$}}}]\addtocounter{CBtnc}{1}jpc@physics.unimelb.edu.au; j%
pcostella@hotmail.com; www.ph.unimelb.edu.au/$\sim$jpc\item[{%
\hfill\makebox[0ex][r]{\raisebox{0ex}[1ex][0ex]{$^{\mbox{$%
\fnsymbol{CBtnc}$}}$}}}]\addtocounter{CBtnc}{1}mckellar@physics.%
unimelb.edu.au\item[{\hfill\makebox[0ex][r]{\raisebox{0ex}[1ex][%
0ex]{$^{\mbox{$\fnsymbol{CBtnc}$}}$}}}]\addtocounter{CBtnc}{1}ar%
awlins@physics.unimelb.edu.au\addtocounter{CBtnc}{1}\item[{\hfill
\makebox[0ex][r]{\raisebox{0ex}[1ex][0ex]{$^{\mbox{$\fnsymbol{CB%
tnc}$}}$}}}]\addtocounter{CBtnc}{1}gjs@baryon.phys.unm.edu; gjs@%
swcp.com\addtocounter{CBcit}{1}\item[\hfill$^{\arabic{CBcit}}$]%
\renewcommand\theCscr{\arabic{CBcit}}\protect\refstepcounter{Csc%
r}\protect\label{cit:Costella1995}J.~P.~Costella\ and B.~H.~J.~M%
cKellar, \renewcommand\theCscr{Costella\ and McKellar}\protect
\refstepcounter{Cscr}\protect\label{au:Costella1995}\renewcommand
\theCscr{1995}\protect\refstepcounter{Cscr}\protect\label{yr:Cos%
tella1995}``The Foldy--Wouthuysen transformation,'' Am. J. Phys.%
\ {\bf63}, 1119--1121\ (1995).\addtocounter{CBcit}{1}\item[\hfill
$^{\arabic{CBcit}}$]\renewcommand\theCscr{\arabic{CBcit}}\protect
\refstepcounter{Cscr}\protect\label{cit:Costella1997}J.~P.~Coste%
lla, B.~H.~J.~McKellar, and A.~A.~Rawlinson, \renewcommand
\theCscr{Costella, McKellar, and Rawlinson}\protect
\refstepcounter{Cscr}\protect\label{au:Costella1997}\renewcommand
\theCscr{1997}\protect\refstepcounter{Cscr}\protect\label{yr:Cos%
tella1997}``Classical antiparticles,'' Am. J. Phys.\ {\bf65}, 83%
5--841\ (1997).\addtocounter{CBcit}{1}\item[\hfill$^{\arabic{CBc%
it}}$]\renewcommand\theCscr{\arabic{CBcit}}\protect
\refstepcounter{Cscr}\protect\label{cit:Pais1982}A.~Pais, 
\renewcommand\theCscr{Pais}\protect\refstepcounter{Cscr}\protect
\label{au:Pais1982}\renewcommand\theCscr{1982}\protect
\refstepcounter{Cscr}\protect\label{yr:Pais1982}\mbox{}\protect
\/{\protect\em Subtle is the Lord\protect\/} (Oxford Univ.\ Pres%
s, Oxford, 1982), p.~$\!$143.\addtocounter{CBcit}{1}\item[\hfill
$^{\arabic{CBcit}}$]\renewcommand\theCscr{\arabic{CBcit}}\protect
\refstepcounter{Cscr}\protect\label{cit:Ungar1991}A.~A.~Ungar, 
\renewcommand\theCscr{Ungar}\protect\refstepcounter{Cscr}\protect
\label{au:Ungar1991}\renewcommand\theCscr{1991}\protect
\refstepcounter{Cscr}\protect\label{yr:Ungar1991}``Thomas preces%
sion and its associated grouplike structure,'' Am. J. Phys.\ {\bf
59}, 824--834\ (1991).\addtocounter{CBcit}{1}\item[\hfill$^{%
\arabic{CBcit}}$]\renewcommand\theCscr{\arabic{CBcit}}\protect
\refstepcounter{Cscr}\protect\label{cit:Jackson1999}J.~D.~Jackso%
n, \renewcommand\theCscr{Jackson}\protect\refstepcounter{Cscr}%
\protect\label{au:Jackson1999}\renewcommand\theCscr{1999}\protect
\refstepcounter{Cscr}\protect\label{yr:Jackson1999}\mbox{}%
\protect\/{\protect\em Classical Electrodynamics\protect\/}, 3rd%
 ~ed. (Wiley, New York, 1999), \mbox{Sec.~$\:\!\!$}11.8.%
\addtocounter{CBcit}{1}\item[\hfill$^{\arabic{CBcit}}$]%
\renewcommand\theCscr{\arabic{CBcit}}\protect\refstepcounter{Csc%
r}\protect\label{cit:Goldstein1980}H.~Goldstein, \renewcommand
\theCscr{Goldstein}\protect\refstepcounter{Cscr}\protect\label{a%
u:Goldstein1980}\renewcommand\theCscr{1980}\protect
\refstepcounter{Cscr}\protect\label{yr:Goldstein1980}\mbox{}%
\protect\/{\protect\em Classical Mechanics\protect\/}, 2nd~ed. (%
Addison-Wesley, Massachusetts, 1980), p.~$\!$285.\addtocounter{C%
Bcit}{1}\item[\hfill$^{\arabic{CBcit}}$]\renewcommand\theCscr{%
\arabic{CBcit}}\protect\refstepcounter{Cscr}\protect\label{cit:M%
acKeown1997}P.~K.~MacKeown, \renewcommand\theCscr{MacKeown}%
\protect\refstepcounter{Cscr}\protect\label{au:MacKeown1997}%
\renewcommand\theCscr{1997}\protect\refstepcounter{Cscr}\protect
\label{yr:MacKeown1997}``Question \#57. Thomas precession,'' Am. 
J. Phys.\ {\bf65}, 105\ (1997).\addtocounter{CBcit}{1} \item[%
\hfill$^{\arabic{CBcit}}$]\renewcommand\theCscr{\arabic{CBcit}}%
\protect\refstepcounter{Cscr}\protect\label{cit:Goedecke1978}G.~%
H.~Goedecke, \renewcommand\theCscr{Goedecke}\protect
\refstepcounter{Cscr}\protect\label{au:Goedecke1978}\renewcommand
\theCscr{1978}\protect\refstepcounter{Cscr}\protect\label{yr:Goe%
decke1978}``Geometry of the Thomas precession,'' Am. J. Phys.\ {%
\bf46}, 1055--1056\ (1978).\addtocounter{CBcit}{1}\item[\hfill$^%
{\arabic{CBcit}}$]\renewcommand\theCscr{\arabic{CBcit}}\protect
\refstepcounter{Cscr}\protect\label{cit:Muller1992}R.~A.~Muller, 
\renewcommand\theCscr{Muller}\protect\refstepcounter{Cscr}%
\protect\label{au:Muller1992}\renewcommand\theCscr{1992}\protect
\refstepcounter{Cscr}\protect\label{yr:Muller1992} ``Thomas prec%
ession: Where is the torque?''\mbox{$\protect\displaystyle\:$}Am%
. J. Phys.\ {\bf60}, 313--317\ (1992).\addtocounter{CBcit}{1} 
\item[\hfill$^{\arabic{CBcit}}$]\renewcommand\theCscr{\arabic{CB%
cit}}\protect\refstepcounter{Cscr}\protect\label{cit:Philpott199%
6}R.~J.~Philpott, \renewcommand\theCscr{Philpott}\protect
\refstepcounter{Cscr}\protect\label{au:Philpott1996}\renewcommand
\theCscr{1996}\protect\refstepcounter{Cscr}\protect\label{yr:Phi%
lpott1996}``Thomas precession and the Li\'enard--Wiechert field,%
'' Am. J. Phys.\ {\bf64}, 552--556\ (1996).\addtocounter{CBcit}{%
1}\item[\hfill$^{\arabic{CBcit}}$]\renewcommand\theCscr{\arabic{%
CBcit}}\protect\refstepcounter{Cscr}\protect\label{cit:Mandl1984%
}F.~Mandl\ and G.~Shaw, \renewcommand\theCscr{Mandl\ and Shaw}%
\protect\refstepcounter{Cscr}\protect\label{au:Mandl1984}%
\renewcommand\theCscr{1984}\protect\refstepcounter{Cscr}\protect
\label{yr:Mandl1984}\mbox{}\protect\/{\protect\em Quantum Field 
Theory\protect\/} (Wiley, Chichester, 1984), \mbox{Secs.~$\:\!\!%
$}8.7 and A.8.\addtocounter{CBcit}{1}\item[\hfill$^{\arabic{CBci%
t}}$]\renewcommand\theCscr{\arabic{CBcit}}\protect\refstepcounter
{Cscr}\protect\label{cit:Misner1970}C.~W.~Misner, K.~S.~Thorne, 
and J.~A.~Wheeler, \renewcommand\theCscr{Misner, Thorne, and Whe%
eler}\protect\refstepcounter{Cscr}\protect\label{au:Misner1970}%
\renewcommand\theCscr{1970}\protect\refstepcounter{Cscr}\protect
\label{yr:Misner1970}\mbox{}\protect\/{\protect\em Gravitation%
\protect\/} (Freeman, New York, 1970).\renewcommand\theCscr{%
\arabic{CBcit}}\protect\refstepcounter{Cscr}\protect\label{citla%
st}\settowidth\Cscr{$^{\ref{cit:Misner1970}}$}\end{list}\par\end
{document}